# Mid-infrared emission and absorption in strained and relaxed direct band gap GeSn semiconductors

S. Assali,[1,+,*] A. Dijkstra,[2,+] A. Attiaoui,[1] É. Bouthillier,[1] J. E. M. Haverkort,[2] and O. Moutanabbir[1]

[1] Department of Engineering Physics, École Polytechnique de Montréal, C. P. 6079, Succ. Centre-Ville, Montréal, Québec H3C 3A7, Canada

[2] Department of Applied Physics, Eindhoven University of Technology, 5600 MB Eindhoven, The Netherlands

By independently engineering strain and composition, this work demonstrates and investigates direct band gap emission in the mid-infrared range from GeSn layers grown on silicon. We extend the room-temperature emission wavelength above ~4.0 μm upon post-growth strain relaxation in layers with uniform Sn content of 17 at.%. The fundamental mechanisms governing the optical emission are discussed based on temperature-dependent photoluminescence, absorption measurements, and theoretical simulations. Regardless of strain and composition, these analyses confirm that single-peak emission is always observed in the probed temperature range of 4-300 K, ruling out defect- and impurity-related emission. Moreover, carrier losses into thermally-activated non-radiative recombination channels are found to be greatly minimized as a result of strain relaxation. Absorption measurements validate the direct band gap in strained and relaxed samples at energies closely matching photoluminescence data. These results highlight the strong potential of GeSn semiconductors as versatile building blocks for scalable, compact, and silicon-compatible mid-infrared photonics and quantum opto-electronics.

# I. INTRODUCTION

Free-space communications, infrared harvesting, biological and chemical sensing, and imaging technologies would strongly benefit from the availability of scalable, cost-effective, and silicon- (Si-) compatible mid-infrared (MIR) opto-electronic devices. With this perspective, Sn-containing group IV semiconductors (Si)GeSn grown on Si wafers have recently been the subject of extensive investigations. [1–5] At the core of these expended efforts is the ability to harness the efficient direct bandgap emission from these emerging semiconductors, which can be achieved at a Sn content around 10 at.% in fully relaxed layers. It is, however, noteworthy that this critical composition is significantly above the ~1 at.% equilibrium solubility of Sn in Ge, which calls for a precise control of the growth kinetics to prevent phase separation and avoid Sn segregation and material degradation. Moreover, the epitaxial growth of GeSn is typically achieved on Si wafers using a Ge interlayer, commonly known as Ge virtual substrate (Ge-VS), resulting in inherently compressively strained GeSn. [6,7] This residual compressive strain increases Sn content threshold for the indirect-to-direct bandgap crossover and the associated optical emission is shifted to higher energies relative to that of a relaxed material. [8–10] Moreover, the accumulated strain in GeSn epitaxial layers was found to affect the incorporation of Sn throughout the growth, which can lead to graded composition as layers grow thicker. [6,7]

Among the strategies to circumvent GeSn growth hurdles, lattice parameter engineering using multi-layer and step-graded growth has been shown to be effective in controlling the Sn content and its uniformity. This process was exploited in recent studies demonstrating room temperature optical emission down to 0.36 eV (~3.4 µm wavelength), [4] as well as optically- and electrically-pumped lasers at short-wave IR (SWIR)-MIR wavelengths operating at lower temperatures. [11–13] Post-growth control and manipulation of strain have also been utilized to extend the emission range as longer emission wavelengths can be achieved through relaxation and tensile strain engineering. [2–4,11,14–17]

Notwithstanding contributions from the aforementioned studies, in-depth investigations of GeSn optical emission are, however, still in their infancy. Besides the ability to engineer innovative optoelectronic devices, the availability of high-quality GeSn layers also provides a rich playground to explore and uncover their fundamental properties and probe their behavior as a function of key parameters, namely temperature, strain, and composition. As mentioned above, strain and composition are always interrelated in epitaxial GeSn layers. Here, we decouple and elucidate their individual effects on the optical emission of GeSn by performing systematic studies on the optical emission of strained and relaxed GeSn in the 4-300 K range. The emission spectra are then discussed in the light of 8×8 $\boldsymbol{k \cdot p}$ band structure calculations and absorption measurements. We found that the optical emission wavelength at room temperature (RT) can be extended above 4 µm upon significant post-growth relaxation of the compressive strain in under-etched $Ge_{0.83}Sn_{0.17}$. By cooling down to 4 K, the single emission peak is preserved in both strained and relaxed layers, indicative of the direct band gap emission across the entire temperature range. Thermally activated non-radiative recombination channels were found to have a negligible effect on the emission of $Ge_{0.83}Sn_{0.17}$ and no additional impurity-related emission were observed upon relaxation. Interestingly, in relaxed $Ge_{0.83}Sn_{0.17}$, the reduction in PL intensity at 300 K is relatively limited and the measured intensity is ~10 % of that recorded at 4 K. Similar behavior was found in lower Sn content layers $Ge_{0.863}Sn_{0.137}$, with an initial in-plane compressive strain of -0.4 %. By performing absorption measurements, the contribution of layers with different composition can be precisely decoupled. These measurements confirm the recorded behavior of PL emission energy upon lattice relaxation, thus providing an additional demonstration of the band gap directness in both strained and relaxed $Ge_{0.83}Sn_{0.17}$ samples, as well in the relaxed $Ge_{0.863}Sn_{0.137}$.

**II. METHODS**

The layers investigated in this work were grown on 4-inch Si (100) wafers in a low-pressure chemical vapor deposition (CVD) reactor using ultra-pure $H_2$ carrier gas, and 10 % monogermane ($GeH_4$) and tin-tetrachloride ($SnCl_4$) precursors, following a recently developed growth protocol. [4,7,18,19] First, a 600-700 nm-thick Ge-VS was grown at 450 °C, followed by thermal cyclic annealing (>800 °C) and additional Ge deposition. The $Ge_{0.83}Sn_{0.17}$ sample with a uniform composition was grown using a GeSn multi-layer heterostructure, where the incorporation of Sn was controlled by the growth temperature. [18] The 17 at.% top layer (TL), 12-10 at.% middle layer (ML), and 8 at.% bottom layer (BL) were grown at 280 °C, 300 °C, and 320 °C, respectively, with a TL/ML/BL thickness of 160/155/65 nm. [4] In the compositionally graded $Ge_{0.863}Sn_{0.137}$ sample, a 8 at.% buffer layer (320 °C) with a thickness of ~60 nm was used for the subsequent growth of the (graded-uniform) 10.5/12.9-13.7 at.% layer at 300 °C on top. [19] To relax the epitaxial strain, the grown layers were released from the substrates by patterning and underetching micro-disks. First, 7 μm-wide disks were defined in ma-N 2410 resist with electron beam lithography. Micro-disks were patterned with a 10 μm pitch over 2.5×2.5 $mm^2$ arrays. This was followed by two successive reactive ion etching (RIE) steps. First, a $Cl_2$-based plasma was used to vertically etch the GeSn layer and transfer the resist pattern to the Ge-VS. Second, Ge was selectively etched in a $CF_4$-based plasma, resulting in released GeSn micro-disks.

The luminescence properties of the samples were measured using a Fourier transform infrared spectrometry (FTIR) based photoluminescence setup. [4] The investigated samples were mounted in a vertically oriented helium flow cryostat with accurate temperature control. The samples were then illuminated using a 35 kHz modulated CW 976 nm laser to allow for lock-in technique. The laser was focused using a 20.3 mm focal length, off-axis parabolic mirror which also collected the PL signal and coupled it through the FTIR system which was equipped with a HgCdTe detector and a Potassium Bromide beam splitter. The full path of the PL emission was nitrogen purged to prevent water absorption lines in the measured spectra. To filter out the laser from the PL signal a germanium window was used as a long-pass filter. The PL measurement on the as-grown samples were performed at an excitation density of ~1 kW/cm$^2$, while a higher value of 4 kW/cm$^2$ was used for the micro-disks. To derive the Arrhenius parameters in a comparable way, all samples have been measured at 4 kW/cm$^2$. The absorption spectra were measured by coupling a light source through a Thermo Scientific Nicolet IS50R FTIR system, focusing it through a sample into a gold-coated integrating sphere and collected at a baffled port using a liquid nitrogen cooled HgCdTe detector. As a light source a LEUKOS ELECTRO MIR 4.1 supercontinuum laser has been used for the as-grown $Ge_{0.83}Sn_{0.17}$ and $Ge_{0.863}Sn_{0.137}$ samples and a glowbar for the micro-disk etched $Ge_{0.83}Sn_{0.17}$ sample because the absorption edge shifted out of the spectrum of the supercontinuum. The full optical path could not be purged with N$_2$ therefore background spectra have been measured directly before every transmission measurement. A schematic of the used setup for this experiment is shown in Supplemental Material S1.

## III. RESULTS AND DISCUSSION

### A. Fabrication and band structure of highly relaxed Ge$_{0.83}$Sn$_{0.17}$

Fig. 1a exhibits a cross-sectional scanning transmission electron microscopy (STEM) image of the GeSn 17/12/8 at.% (TL/ML/BL) multi-layer stacking, with thicknesses of 160/155/65 nm, [4] grown on Ge-VS/Si substrate. The composition of each layer is estimated from Reciprocal Space Mapping (RSM) around the asymmetrical (224) X-ray diffraction (XRD) peak (Fig. 1b). We note that the Sn content values obtained from the RSM map can be underestimated by 0.5-1.0 at.% as compared to values obtained using atom probe tomography measurements. [4] This difference may come from the difference in the volumes analyzed by each technique. For consistency, compositions extracted from RSM will be employed as they are averaged over a volume comparable to that probed in PL studies. The RSM data indicate that the TL is under an in-plane biaxial compressive strain $\varepsilon_{||} = -1.3$ %. As mentioned in the previous section, to significantly relax this residual strain, it suffices to release the GeSn layers by etching away the underlying Ge VS, as illustrated in Fig. 1c. The scanning electron microscope (SEM) image in Fig. 1d shows a typical array of micro-disks having a diameter of 7 μm and a pitch of 10 μm. The CF$_4$ etching time was selected to completely release the GeSn micro-disks which collapsed onto the Si wafer. We note that a small residual thickness of the Ge-VS layer is visible on the Si substrate underneath the micro-disk after the fabrication process (Supplemental Material S2), which does not affect the strain in the micro-disks because they are detached from the substrate.

Raman measurements were performed on the micro-disk arrays to evaluate the residual strain in the GeSn TL. Note that Raman spectra are recorded from the TL without any contribution from the underlaying layers as the penetration depth of the 633 nm excitation laser (<30 nm) is significantly smaller than the TL thickness (160 nm). [4] Fig. 1e displays Raman spectra acquired at the center of a single micro-disk (red curve). As a reference, the Raman spectrum of the as-grown $Ge_{0.83}Sn_{0.17}$ layer is also shown (blue curve). The Ge-Ge LO mode of the as-grown layer is centered at ~292 cm$^{-1}$, whereas the same mode shifts down to ~287 cm$^{-1}$ in the under-etched micro-disks. The observed ~5 cm$^{-1}$ shift corresponds to a strain relaxation from the as-grown value of -1.3 % down to -0.2 % in the micro-disks. Note that Raman spectra associated to the area between micro-disks only show the Si-Si LO mode at 520 cm$^{-1}$ from the substrate (Supplemental Material S2), indicating that the Ge-VS was completely etched leading to the observed strain relaxation. The measured residual strain (-0.2 %) in the TL even after the complete release of the micro-disks is expected because the GeSn stack includes three layers with a variable composition. This post-etching strain analysis is consistent with systematic studies decoupling Sn content and strain effects on GeSn Raman vibrational modes. [18]

To elucidate the effects of composition and strain on the optical properties of the investigated GeSn stacking, an accurate theoretical framework to evaluate the band structure of these multilayers is highly coveted. For instance, to be able to interpret any optical transition, the band gap in each layer of the TL/ML/BL stacking needs to be estimated as a function of strain and Sn content. To that end, we employed the 8-band $\mathbf{k \cdot p}$ model framework at 300 K together with the envelope function approximation (EFA). [20] The 8-band $\mathbf{k \cdot p}$ GeSn parametrization was based on the work of Chang *et al*, [21] while strain implementation is based on the Bir-Pikus formalism. [22] Additionally, the model solid theory was adopted to estimate the conduction and valence band offset profiles of various high symmetry bands. [23,24] Note that Vegards's law is inaccurate to estimate the bandgaps of GeSn compounds due to significant bowing effects of the $L$ and $\Gamma$ high symmetry points. To circumvent this limitation, a new set of temperature-dependent bowing parameters at the Γ direction was extracted from our temperature-dependent PL measurement for the strained and relaxed 17 at.% TL as well as the relaxed 13.7 at.% ML. More detail about the undertaken fitting procedure is highlighted in the Supplemental Material S3. In a nutshell, at 300K, the band gap bowing parameters $b_\Gamma^{strained-TL}$ and $b_\Gamma^{relaxed-TL}$ are estimated to be 2.18 eV and 2.52 eV, respectively whereas $b_\Gamma^{relaxed-ML}$ is equal to 2.27 eV. The newly extracted Γ-L direction band gap bowing parameter matches well with values established in literature. [25,26] The obtained band lineup diagrams for the $Ge_{0.92}Sn_{0.08}/Ge_{0.88}Sn_{0.12}/Ge_{0.83}Sn_{0.17}$ stack are shown in Fig. 2 for the as-grown and relaxed layers at 300 K. From these diagrams, the electron and holes are predicted to diffuse into the upmost 17 at. % layer, as a result of the band offset with the 12 at.% ML being equal to 27 and 65 meV for both electrons and heavy holes (HH), respectively. Besides, the strain relaxation (from -1.3% to -0.2%) in the TL layer induces an increase in the energy difference between $L$ and Γ minima in the conduction band $E_g^{L-\Gamma}$ ($= E_g^L - E_g^\Gamma$) of 19 meV, which should enhance band-to-band recombination. Furthermore, the LH-HH splitting is reduced by 82 meV in the TL layer. The valence (conduction) band offset between the ML and TL is reduced (increased) by 18 meV (45 meV) due to strain relaxation. Note that Low *et al.* [27] have shown using empirical pseudopotential methods that the band gaps for the unstrained $Ge_{0.83}Sn_{0.17}$ at Γ direction to be equal to 0.42 eV (at 4 K). This compares well with our ***k·p*** parametrization which gives 0.39 eV. Additionally, based on Gupta *et al.*, [8] the as-grown ML layer (12% at. Sn, -0.54% compressive strain) is expected to have a direct band gap of 0.51 eV (at 0 K), which is in perfect agreement with our ***k·p*** model indicating that the ML direct gap transition equals 0.50 eV (at 0 K). Our calculations predict a type-I band alignment in the GeSn multi-layer heterostructure, where the radiative recombination should occur in the 17 at.% TL. These calculated results are discussed further below in the light of the optical measurements.

**B. Room temperature PL and transmission measurements for Ge$_{0.83}$Sn$_{0.17}$**

To experimentally investigate the effect of strain relaxation on the optical properties of GeSn, room-temperature PL and transmission measurements were performed on both as-grown layers and etched micro-disk arrays. We first focus on the PL spectra displayed in Fig. 3. In the as-grown Ge$_{0.83}$Sn$_{0.17}$, a main emission peak centered at 0.365±0.005 eV (*i.e.* 3.4 µm wavelength) is recorded with a full-width at half maximum (FWHM) of 40-50 meV. A low intensity emission peak at ~0.43 eV is also visible, which could be related either to the optical recombination in the underlying 12 at.% ML or to optical transitions involving the LH-band instead of the HH-band, as will be discussed later in the context of data displayed in Fig. 4. Note that the optical emission in this sample mainly originates from the 17 at.% TL. [4] In the micro-disks, the optical emission is redshifted to 0.315±0.005 eV (FWHM~60-70 meV) compared to the as-grown sample, thus covering a broader range in the MIR up to 4.0-4.5 µm (Fig. 3b). The measured ~50 meV shift (*i.e.,* ~0.5 µm in wavelength) in the optical emission is induced by the strain relaxation from the as-grown value of -1.3 % to -0.2 % in the micro-disks. The observed increase in the FWHM of the emission peak in the micro-disks could be related to small fluctuations in strain between different micro-disks, or due to strain-induced HH-LH band splitting decreasing from ~100 meV to ~20 meV in the micro-disks, which would make the LH also contribute with a new emission peak, partially overlapping with the HH peak and thus broadening the PL spectrum. A comparison with published data must take in consideration both composition and strain values to reach an accurate assessment. Von Driesch *et al.* showed PL emission down to 0.39 eV in partially-relaxed Ge$_{0.86}$Sn$_{0.14}$ ($\varepsilon_{||}$=-0.6 %), [14] and Calvo *et al.* demonstrated PL emission at ~0.37 eV in Ge$_{0.84}$Sn$_{0.16}$ ($\varepsilon_{||}$~-0.8 %). [3] These results are consistent with the lower emission energies observed for our strained and relaxed Ge$_{0.83}$Sn$_{0.17}$ layers, as displayed in Fig. 3. However, the reported optical emission at 0.385 eV in significantly relaxed $Ge_{0.825}Sn_{0.175}$ ($\varepsilon_{||}$~-0.3 %) [2] deviates from this trend and seems to be higher than what it should be. In fact, at nearly the same composition and strain in $Ge_{0.83}Sn_{0.17}$ micro-disks the emission energy is 0.31 eV (Fig. 3).

The optical emission from $Ge_{0.83}Sn_{0.17}$ is then correlated with the absorption measurements performed at 300 K, as shown in Supplemental Material S4-6. To provide a precise evaluation of the transmission data and to be able to determine the band gaps of individual layers we define the Napierian absorbance as:

$$A_e = \sum_i \alpha_i \cdot d_i / \cos(\theta_i), \quad (1)$$

where $\alpha_i$ and $d_i$ are the absorption coefficient and thickness of layer $i$ in the stack respectively and $\theta_i$ the angle with respect to the normal under which the light travels through the layer in the transmission experiment. Surface reflections and interference effects have been minimized (see Fig S6) by performing transmission experiments at the Brewster angle in an integrating sphere for the as-grown sample as elaborated in the Supplemental Material S4-6. The micro-disks showed not to be affected by interference fringes, probably because of misorientation of the individual disks and a slight surface roughness due to etching, therefore they could be measured at normal incidence. The obtained $A_e^2$ curves are plotted in Fig. 3 together with their respective PL spectra. The $A_e^2$ spectrum for the as-grown sample clearly shows two onsets that are associated with the TL and ML in the sample where the disks show a single onset associated with the relaxed TL. In a direct band gap semiconductor, the absorption coefficient $\alpha$ scales as the square root of the energy and linearly with $A_e$. Therefore, $A_e^2$ shows a linear behavior with energy, where the band gap is given by the energy-axis crossing. The band gaps determined using this method are also indicated in Fig. 3 as dashed vertical lines. A bandgap for the as-grown 17 at.% ($\varepsilon_{||} = -1.3$ %) sample of 0.35±0.01 eV is found, which lies on the rising edge of the corresponding PL signal centered at 0.365 eV as expected. Only a limited contribution from an Urbach tail is visible in the $A_e^2$ spectrum, in agreement with a previous report on GeSn with a composition up to 10 at.% Sn. [28] Considering the temperature dependence of PL emission (*i.e.* direct band gap), as discussed later in Figs. 4-5, a band gap of ~0.29 eV is predicted using the 8-band $\mathbf{k} \cdot \mathbf{p}$ method at 300 K (see Supplemental Material S3). A small difference between the PL and $\mathbf{k} \cdot \mathbf{p}$ values could arise from the use of a non-optimized deformation potential constant, due to the scarcity of accurate experimental studies available in literature. In this work, the deformation potential constants were based on the work of Chang *et al.* [21]. Similarly, the bandgap of the 12 at.% Sn ($\varepsilon_\parallel = -0.5$ %) ML is found to be 0.43±0.01 eV. The extrapolated 8-band $\mathbf{k} \cdot \mathbf{p}$ band gap value at 300 K is in the 0.34-0.36 eV range, depending on the strain value considered (see Supplemental Material S3). The 0.43 eV edge is in close agreement with the ~0.45 eV estimated for Ge$_{0.875}$Sn$_{0.125}$ ($\varepsilon_\parallel = -0.3$ %) using reflection measurements by Driesch *et al.*. [14] Nonetheless, the values for both samples agree closely and show the robustness of the method.

In the Ge$_{0.83}$Sn$_{0.17}$ micro-disks only a single absorption onset is visible at 0.31±0.01 eV and a rather flat $A_e^2$ curve is observed above 0.45 eV, without any additional absorption edges at higher energy associated with the 12-8 at.% Sn layers. The absence of additional absorption onsets can plausibly be attributed to the significantly lower signal-to-noise ratio in the transmission measurements obtained for micro-disks because of a reduced absorption volume and the limited intensity of required IR light source. We note that the absorbance of the micro-disks shown in Fig. 3b (in which the filling factor has been taken into account) is similar to the as-grown 17/12 at.% layers (TL/ML) stacking, thus indicating no deterioration of the material quality after the etching process.

## C. Temperature-dependent PL measurements in Ge$_{0.83}$Sn$_{0.17}$

To investigate the evolution of the direct band gap with temperature and strain and to identify any possible optically active defect levels, temperature-dependent PL spectra were acquired in the 4-300 K temperature range for both the as-grown and the micro-disk sample. These results are displayed in Fig. 4a-d. For the as-grown Ge$_{0.83}$Sn$_{0.17}$ layer at 4 K (Fig. 4a), a single emission peak at 0.413±0.001 eV is observed (FWHM=19.0±0.5 meV). The symmetric shape of the peak is attributed to limited band-filling, indicating a low-carrier injection regime, as shown in the power-dependent PL measurements at 4 K in the Supplemental Material S7. The 4 K PL emission at very low excitation power most likely originates from shallow localized states, such as bound-excitons or free-to-bound transitions, as suggested by the constant emission energy with varying excitation power. [29–31] Local fluctuations in Sn content across the sample and the presence of (low concentration) impurities are most likely responsible for the observed behavior at low temperatures. At higher excitation power band-to-band emission dominates with a Burstein-Moss shift of 2-6 meV/decade. [32] Note that no power-dependent energy shift, using similar excitation power densities, was measured at 300 K, [4] which further confirms that the direct band gap is maintained at room-temperature. In addition, the increase in PL intensity remains linear (slope of ~1) even at the highest excitation power densities, indicating that Auger non-radiative recombination does not play a role (see Supplemental Material S7), which is rather surprising for a narrow band gap semiconductor. When the temperature is increased to ~50 K, a small blueshift (1.0-1.5 meV) is observed, followed by a progressive redshift as the temperature increases to 300 K, as displayed in Fig. 4c. No additional peaks are detected in the whole temperature range. The temperature-dependent PL spectra for the Ge$_{0.83}$Sn$_{0.17}$ micro-disk arrays are shown in Fig. 4b. A main emission peak centered at 0.365±0.001 eV (FWHM=42±3 meV) is detected at 4 K. The emission energy remains constant up to 50 K, followed by a progressive redshift at higher temperatures, reaching 0.315 eV at 300 K. We note that the absence of a small blueshift (<2 meV) below 50 K might simply result from the higher FWHM in micro-disks compared to as-grown layers. We highlight that the ~50 meV energy difference recorded at 50 K between the as-grown layer and the micro-disks is close to the 45 meV observed at 300 K. Thus, the strain in the multilayer structure changes only by a negligible amount with temperature without affecting the band gap directness across the probed temperature range. The absence of additional PL peaks indicates that the optical recombination dominates in the $Ge_{0.83}Sn_{0.17}$ layer regardless of temperature.

To estimate the temperature-dependent band gap from the PL measurements a line-shape model is fitted to the spectra shown in Fig. 4a-b according to the equation [33,34]:

$$I_{PL} = A * \left[ \sqrt{E - E_g} \cdot \exp\left(-\frac{E}{k_B T}\right) \right] * \left[ \frac{1}{\gamma \sqrt{2\pi}} \exp\left(-\frac{E^2}{2\gamma^2}\right) \right] \qquad (2),$$

where $A$ is a scaling constant, T is the temperature, and $\gamma$ is a broadening factor. The first term of the model is given by the joint density-of-states (JDOS) multiplied by a Boltzmann distribution, assuming low-excitation conditions and parabolic bands. In the second term, broadening mechanisms, such as alloy broadening and strain fluctuations, are taken into account by the convolution of the initial line-shape with a Gaussian distribution. We note that the best results for the fitting procedure were obtained by setting the carrier and lattice temperatures to be identical, as discussed in more detail in Supplemental Material S8. The temperature dependence of the PL peak and direct band gap energies of $Ge_{0.83}Sn_{0.17}$ are displayed in Fig. 4c. In both as-grown and relaxed samples, the direct band gap is 10-20 meV lower in energy compared to the PL peak across the 4-300 K temperature range. At room temperature, the band gaps of the as-grown sample (0.34 eV) and the micro-disks sample (0.29 eV) are only slightly below the values determined using absorption measurements (0.35±0.01 eV and 0.31±0.01 eV respectively) shown in Fig. 3. This very small shift is probably a result of alloying effects, where PL is more likely to probe the lowest band gap material and the absorption measurement averages all contributions. Temperature-independent broadening factors $\gamma$ were obtained for both the as-grown ($\gamma = 13 \pm 1$ meV) and the micro-disk sample ($\gamma = 25 \pm 2$ meV), as plotted in the inset of Fig. 4c. The increased broadening in the micro-disks with respect to the as-grown sample can be attributed to small strain fluctuations that are introduced in the etching process. Next, the fit of the band gap as a function of temperature is performed using the Vina's equation: [35]

$$E_g = a - b\left(1 + \frac{2}{\exp\left(\frac{\theta_D}{T}\right)-1}\right), \qquad (3)$$

where $a$ is a constant, $b$ is the strength of the electron-phonon interaction, and $\theta_D$ is the mean temperature of the phonons contributing to the scattering process. The values obtained using Eq. (3) are listed in Table 1. When compared to the widely used Varshi's equation, [36] the Vina's equation provides higher fit accuracy in the high temperature regime for materials with lower $\theta_D$ like Ge (374 K) and α-Sn (270 K) as compared to Si (640 K). [35,37]

The integrated PL intensities, as a function of the inverse of temperature, for the direct band gap of the as-grown $Ge_{0.83}Sn_{0.17}$ layer and the micro-disk array are displayed in Fig. 4d. The data are normalized to the intensity measured at 4 K for each set of samples. Overall, the PL emission exhibits the same qualitative behavior for both samples, *viz.*, a monotonous decrease as temperature increases, which becomes more pronounced above 100 K. At 300 K, a decrease in the PL intensity to 1-2 % compared to the 4 K intensity is measured for the as-grown layer, whereas in the micro-disk array this decrease is rather limited to 10-20 % suggesting an increase in the quantum efficiency in strain relaxed micro-disks. Since no sudden changes in both PL emission energy and intensity are observed across the 4-300 K range, direct band gap radiative recombination is the dominating mechanism in the as-grown and the micro-disks samples, [38,39] with a negligible amount of impurity ionization. [40]

To shed more light on the non-radiative processes in these samples, the integrated PL intensities are fitted with Arrhenius functions (dashed lines in Fig. 4d) considering two active non-radiative recombination channels, according to:

$$I = 1/\left[1 + c_1 \cdot \exp\left(-\frac{E_a^1}{kT}\right) + c_2 \cdot \exp\left(-\frac{E_a^2}{kT}\right)\right], \quad (4)$$

where $c_1$ and $c_2$ are constants, $E_a^1$ and $E_a^2$ are the activation energies, and $k$ is the Boltzmann constant. [41] This set of parameters contains rich information on physical recombination processes. [39] The results obtained from the fit using Eq. (4) are listed in Table 2. The first activation energy $E_a^1$ is ~5 meV in as-grown layers and micro-disks suggesting a common loss mechanism for the two materials. This loss could be facilitated by a shallow impurity level ionizing at low temperatures, perhaps related to the observed 1-2 meV blueshift in as-grown layer at low temperatures. Thus, in Ge$_{0.83}$Sn$_{0.17}$ the recorded PL below 50 K is most likely due to a 1-4 meV shallow level, whereas band-to-band recombination is observed at higher temperatures. [31,40] The second activation energy $E_a^2$ indicates the presence of a 20-40 meV higher energy non-radiative recombination channel, which could originate from inter-valley tunneling of electrons into the L-minimum or from a deeper impurity level within the band gap. The calculated band diagram of Ge$_{0.83}$Sn$_{0.17}$ (Fig. 2) indicates that the L-minimum is ~103 meV and ~122 meV higher compared to the Γ-minimum in as-grown (strained) and micro-disks (relaxed) samples, respectively. Thus, inter-valley tunneling is an unlikely carrier diffusion process in Ge$_{0.83}$Sn$_{0.17}$ across the entire 4-300 K temperature range. Other possible loss mechanisms are surface traps or defects within the layer (*i.e.*, impurities and point defects), which are both thermally activated sources of non-radiative recombination. Herein, it is important to mention that no impurities are observed in atom-probe tomography (APT) measurements (detection limit ~1·10$^{17}$ cm$^{-3}$) [42] of Ge$_{0.83}$Sn$_{0.17}$ layer (Supplemental Material S9) [4]. However, point defects in a semiconductor, such as vacancies and vacancy complexes, can also affect the optical properties due to carrier trapping. [43–45] The presence of divacancies and, to a limited extent, of vacancy-clusters at room-temperature has been recently confirmed in as-grown GeSn layers [19] and they might contribute to the activation energy $E_a^2$ as an additional non-radiative recombination channel. Another plausible interpretation for $E_a^2$ would be the loss of carriers due to diffusion from the 17 at.% TL to the 12 at.% ML, where they (partially) recombine non-radiatively. In the as-grown sample, the calculated conduction band offset between TL and ML, $\Delta E_c$=27 meV, (Fig. 2) is close to the activation energy $E_a^2$ (32±7 meV) as provided by the fit. This means that electrons can diffuse to the ML, where an overall quench in PL will already be observed when only one of the two charge carriers leaks away. With this mechanism in mind, one might argue that the observed ~0.45 eV PL peak in the as-grown Ge$_{0.83}$Sn$_{0.17}$ above 180 K (Fig. 4a) sample could be associated with emission from the 12 at.% ML instead of a contribution of LH-HH splitting in the strained 17 at.% TL. In the relaxed ML layer, the conduction and valence band offsets (at 0 K), respectively equal to $\Delta E_c$=78 meV and $\Delta E_v$=30 meV) changes such that only holes can be thermally excited into the ML.

**D. Effect of composition on the optical properties.**

In addition to the post-growth relaxation described above and for the sake of comparison, we examine in the following strain relaxation achieved during epitaxial growth using a compositionally graded GeSn layer. In this configuration, the lattice parameter gradually increases throughout the layer thickness and additional Sn atoms are incorporated in the growing layer, which results in a graded profile. [7] To put the effect of strain relaxation and composition in perspective, we compare the results in Figs. 3-4 with a lower Sn-content graded $Ge_{0.863}Sn_{0.137}$ sample that is relatively highly relaxed ($\varepsilon_{||} = -0.4\ \%$) and shows PL emission at a similar energy to that of as-grown $Ge_{0.83}Sn_{0.17}$ sample. The high relaxation on the other hand makes a direct comparison with the previously discussed micro-disks possible while considering the difference in composition. The compositionally graded growth starts with a ~60 nm-thick nucleation layer with 8 at.% Sn, followed by a ~700 nm-thick GeSn layer (Fig. 5a). In this layer, a graded composition (60-400 nm) from 10.5 to 12.9 at.% is observed, followed by a uniform composition (400-700 nm) of 13.7 at.%. The graded and uniform layers are also visible in the RSM map in Fig. 5b, as described in references [7,19]. The residual in-plane strain across the GeSn layers is lower than -0.4 %, thus indicating a high degree of strain relaxation. Room-temperature PL spectra measured for this sample are plotted in Fig. 5c. A main emission peak centered at 0.380±0.010 eV is observed, however, with a relatively large FWHM of ~100 meV that most likely results from the gradual change in the band gap energy across the graded layer (*i.e.* emission from the lower Sn content layers). Note that in the as-grown $Ge_{0.83}Sn_{0.17}$ sample with a uniform composition, the FWHM is half of the value obtained in the compositionally graded $Ge_{0.863}Sn_{0.137}$. Additionally, the 0.38 eV emission in $Ge_{0.863}Sn_{0.137}$ is in line with the emission at ~0.39 eV reported for partially relaxed Ge$_{0.86}$Sn$_{0.14}$ ($\varepsilon_{||} = -0.6$ %) [14] and in lower Sn content Ge$_{0.893}$Sn$_{0.107}$ micro-disks with a 0.45 % tensile strain induced by external SiN stressors. [15]

In the transmission data and corresponding Napierian absorbance squared curve (Fig. 5c), the direct band gap absorption edge for the layer with a composition of 13.7 at.% is visible at 0.39±0.01 eV, while no other onsets are visible due to the graded at.% Sn composition. The very good agreement between the PL and absorption data demonstrates that the band-to-band radiative recombination takes place in the upmost Ge$_{0.863}$Sn$_{0.137}$ layer. It is important to note that no reliable line shape fitting procedure could be performed because of the graded composition in this sample resulting in a gradual change in the band gap energy. Therefore, the PL peak position was used in the following data analysis. In temperature-dependent PL measurements (Fig. 5d-f), a single emission peak is observed throughout the 4-300 K range. At 4 K, the emission peak is centered at 0.416±0.001 eV (FWHM=42.0±0.5 meV) and its width is 2× larger than that of the Ge$_{0.83}$Sn$_{0.17}$ layer. This difference in FWHM remains similar to that observed at room-temperature. When the temperature is increased the emission energy in Ge$_{0.863}$Sn$_{0.137}$ remains unaffected up to 80 K, followed by a progressive decrease at higher temperature, reaching 0.381 eV at 300 K. When the temperature rises from 4 K to 300 K, the integrated PL intensity drops to ~15 % of its initial value, which is very similar to what is observed in relaxed Ge$_{0.83}$Sn$_{0.17}$ micro-disks (Fig. 4d). The values estimated from the Vina's fit (eq. 3, Table 1) and activation energies $E_a^1$ and $E_a^2$ values from the Arrhenius plot (eq. 4, Table 2) are in the same range as those obtained for Ge$_{0.83}$Sn$_{0.17}$. Therefore, the graded composition in the as-grown Ge$_{0.863}$Sn$_{0.137}$ has a negligible effect on the energy of shallow levels, while it increases the emission linewidth by a factor 2 at all temperatures when compared to Ge$_{0.83}$Sn$_{0.17}$. Despite the broader PL peak, the strain relaxation resulting from the graded composition might promote optical recombination and minimize non-radiative recombination channels, in analogy to the $Ge_{0.83}Sn_{0.17}$ micro-disk arrays discussed above where a similar PL quenching to ~10 % of its initial value was observed from 4 K to 300 K (Fig. 4d). Thus, the optical properties of GeSn are enhanced by strain relaxation, regardless of the method used to reduce strain in the epilayer.

Curiously enough, the negligible role of defect- and impurity-related emission seen in the data exhibited in Fig. 4-5 remains one of the striking observations when the current work is compared to literature despite the higher Sn content in the investigated layers. In fact, the few available studies highlighted clear contributions from dislocations and impurities on the temperature-dependent PL emission in GeSn at Sn content close or above 10 at.%. [38,40,46] For instance, Pezzoli *et al.* discussed the trapping of carriers in dislocations observed at low temperature in indirect band gap $Ge_{0.91}Sn_{0.09}$, indicating that subsequent thermal de-trapping seems to result in an enhanced PL intensity due to increased radiative band-to-band recombination. [38] Defect-related deep levels (~150 meV) were also observed in pseudomorphic (<50 nm-thick) $Ge_{0.875}Sn_{0.125}$ layers, which are possibly associated to the presence of dislocations in GeSn/Ge-VS heterostructure. [46] For thicker layers (>500 nm-thick), the emission from deep-levels was absent, however, a ~8 meV blue-shift of the PL emission energy was clearly observed when temperature increases from 4 K to 100 K ($E_a$~32 meV), followed by a rapid quenching of the emission at higher temperatures. This behavior indicates the presence of shallow impurity levels within the band gap that are ionized above 100 K (shallow level). Impurity ionization (20-30 meV) into the band gap emission with increasing temperature was also observed in $Ge/Ge_{0.87}Sn_{0.13}$ core/shell nanowires [40] and in $Ge_{0.81}Sn_{0.19}$ nanowires. [47] Obviously, any explanation of these differences would be challenging at this stage, and residual impurities are most likely present in our samples with concentrations lower than the detection limit (~1·10$^{17}$ cm$^{-3}$) of APT (Supplemental Material S9). However, it is sensible to consider the influence of growth conditions on the possible incorporation of point defects and impurities. In this regard, one can invoke the growth rate as an obvious difference as the layers investigated in this work are grown at a rate of ~1.5 nm/min, which is ~10 times lower than comparable samples from other research groups. [7] Nonetheless, the growth of GeSn shells around Ge nanowires was performed at a lower growth rate (<1 nm/min) [40] for which PL emission associated with deep levels was observed, thus hinting to other possible critical factors. As a matter of fact, further in-depth investigations are needed to decipher the key parameters affecting the impurity and vacancy incorporation, their complexes, and their role in shaping the optical properties of GeSn semiconductors.

## IV. CONCLUSIONS

In this work, we demonstrated that room-temperature optical emission in GeSn can be extended to longer wavelengths in the mid-infrared range by independently controlling strain and composition. The observed single-peak PL emission at 300 K in the as-grown $Ge_{0.83}Sn_{0.17}$ shifts from 0.365 eV to 0.315 eV (*i.e.* ~3.4 µm to ~4.0 µm wavelength) by releasing most of the -1.3 % compressive strain in under-etched micro-disks, in qualitative agreement with 8×8 ***k·p*** band structure calculations. No additional PL peaks are visible when cooling down as-grown and micro-disk samples to 4 K, suggesting the absence of defect- and impurity-related emission and that direct band gap emission is preserved. By increasing the temperature from 4 K to 300 K, the PL intensity is reduced to 1-2 % and 10-20 % of its initial value in as-grown and relaxed $Ge_{0.83}Sn_{0.17}$, respectively. Therefore, carrier losses into thermally-activated non-radiative recombination channels are minimized by promoting strain relaxation. These observations are confirmed in Ge$_{0.863}$Sn$_{0.137}$, where a significant strain relaxation is obtained as a result of a graded composition. However, a broader 300 K PL emission peak at 0.380 eV is obtained as compared to Ge$_{0.83}$Sn$_{0.17}$ with uniform composition. Regardless of strain and composition, impurities and other non-radiative recombination channels seem to have no detrimental effect on PL emission. Absorption measurements performed at 300 K give a deep insight into the absorption process in GeSn multi-layer structures and confirm the effect of strain relaxation in inducing a redshift of the absorption edge. Additionally, these analyses also demonstrate the band gap directness in strained and relaxed Ge$_{0.83}$Sn$_{0.17}$ layers as well is in the compositionally-graded Ge$_{0.863}$Sn$_{0.137}$. Thus, the control of both strain and composition uniformity is of paramount importance to engineer the emission operational range and linewidth in GeSn opto-electronic devices. [17] Applications requiring a narrower spectral range would benefit from the use of uniform, GeSn layers, where a large amount of Sn can be incorporated while avoiding phase segregation. [48] Whereas, a graded composition would enhance absorption at a larger layer thickness and cover a broader spectral range.

**ACKNOWLEDGMENTS**

The authors thank J. Bouchard for the technical support with the CVD system. O.M. acknowledges support from NSERC Canada (Discovery, SPG, and CRD Grants), Canada Research Chairs, Canada Foundation for Innovation, Mitacs, PRIMA Québec, and Defence Canada (Innovation for Defence Excellence and Security, IDEaS). S.A. acknowledges support from Fonds de recherche du Québec-Nature et technologies (FRQNT, PBEEE scholarship). A.D. acknowledges support from the NWO gravity program.

**AUTHOR INFORMATION**

Corresponding Author:

*E-mail: simone.assali@polymtl.ca

‡: these authors contributed equally to this work.

Notes:

The authors declare no competing financial interest.

**FIGURES CAPTIONS**

**Figure 1.** (a) Cross-sectional TEM image along the [110] zone axis of the GeSn 17/12/8 at.% (TL/ML/BL) multi-layer stacking grown on the Ge-VS/Si substrate. Adapted from Ref. [4]. (b) XRD-RSM around the asymmetrical (224) reflection for the as-grown $Ge_{0.83}Sn_{0.17}$ sample. Adapted from Ref. [4]. (c) Schematics of the micro-disks fabrication process. (d) SEM image of the $Ge_{0.83}Sn_{0.17}$ micro-disks (45° tilting angle). (e) Normalized Raman spectra for $Ge_{0.83}Sn_{0.17}$ acquired on the as-grown (blue curve) and on the central portion of the micro-disk (red curve).

**Figure 2**. (a-b) Calculated 8×8 ***k·p*** band lineup at 300 K for the $Ge_{0.83}Sn_{0.17}$ with an in-plane biaxial strain $\varepsilon_{||} = -1.3\ \%$ (as-grown) (a) and $\varepsilon_{||} = -0.2\ \%$ (micro-disks) (b).

**Figure 3**. (a-b) PL spectra and the Napierian absorbance squared ($A_e^2$) at 300 K for the $Ge_{0.83}Sn_{0.17}$ as-grown (a) and micro-disk (b) samples. In the $A_e^2$ curves, the band gap of the top layer has been estimated as the interception of the dashed linear fit with the x-axis. For the estimation of the second band gap at 0.43 eV in (a) the contribution of the band gap of the first layers has been subtracted.

**Figure 4**. (a-b) Temperature-dependent PL measurements in the 4-300 K range for the $Ge_{0.83}Sn_{0.17}$ as-grown (a) and the micro-disk (b) samples. Spectra are normalized to their own maximum. The line-shape fit using Eq. (2) for each temperature are overlayed as dashed lines, with the band gap as a vertically dotted line. (c) PL emission energy and direct band gap (BG) energy as a function of the temperature. Inset: broadening parameter $\gamma$ (Eq. (2)) as function of the temperature. (d) Integrated PL intensity, normalized to their respective intensity at 4 K, as a function of the inverse of the temperature extracted from the data shown in (a,b). The fit of the BG curves in (c) was performed using Eq. (3), while the data in (d) were fitted using Eq. (4) (dashed curves).

**Figure 5**. (a) Cross-sectional TEM image along the [110] zone axis of the $Ge_{0.863}Sn_{0.137}$ sample grown on the Ge-VS/Si substrate. Adapted from Ref. [19]. (b) XRD-RSM around the asymmetrical (224) reflection. Adapted from Ref. [19]. (c) PL spectra and Napierian absorbance squared ($A_e^2$) at 300 K, (d) Temperature-dependent PL measurements in the 4-300 K range. (e-f) Emission energy as a function of the temperature (e) and integrated PL intensity as a function of the inverse of the temperature (f) extracted from the (d).

**Table 1**. List of the parameters ($a$, $b$, $\theta_D$) extracted from the Vina's fit (Eq. 3) of Fig. 4c and Fig. 5e.

**Table 2**. List of the parameters ($E_a^1$, $E_a^2$, $c_1$, $c_2$) extracted from the fit considering two non-radiative recombination channels (Eq. 4) of Fig. 4d and Fig. 5f.

# REFERENCES

[1] S. Wirths, R. Geiger, N. von den Driesch, G. Mussler, T. Stoica, S. Mantl, Z. Ikonic, M. Luysberg, S. Chiussi, J. M. Hartmann, H. Sigg, J. Faist, D. Buca, and D. Grützmacher, Nat. Photonics **9**, 88 (2015).

[2] J. Margetis, S. Al-Kabi, W. Du, W. Dou, Y. Zhou, T. Pham, P. Grant, S. Ghetmiri, A. Mosleh, B. Li, J. Liu, G. Sun, R. Soref, J. Tolle, M. Mortazavi, and S.-Q. Yu, ACS Photonics **5**, 827 (2018).

[3] V. Reboud, A. Gassenq, N. Pauc, J. Aubin, L. Milord, Q. M. Thai, M. Bertrand, K. Guilloy, D. Rouchon, J. Rothman, T. Zabel, F. Armand Pilon, H. Sigg, A. Chelnokov, J. M. Hartmann, and V. Calvo, Appl. Phys. Lett. **111**, 092101 (2017).

[4] S. Assali, J. Nicolas, S. Mukherjee, A. Dijkstra, and O. Moutanabbir, Appl. Phys. Lett. **112**, 251903 (2018).

[5] A. Attiaoui, S. Wirth, A.-P. Blanchard-Dionne, M. Meunier, J. M. Hartmann, D. Buca, and O. Moutanabbir, J. Appl. Phys. **123**, 223102 (2018).

[6] J. Margetis, S.-Q. Q. Yu, N. Bhargava, B. Li, W. Du, and J. Tolle, Semicond. Sci. Technol. **32**, 124006 (2017).

[7] S. Assali, J. Nicolas, and O. Moutanabbir, J. Appl. Phys. **125**, 025304 (2019).

[8] S. Gupta, B. Magyari-Köpe, Y. Nishi, and K. C. Saraswat, J. Appl. Phys. **113**, (2013).

[9] A. Attiaoui and O. Moutanabbir, J. Appl. Phys. **116**, 063712 (2014).

[10] M. P. Polak, P. Scharoch, and R. Kudrawiec, J. Phys. D. Appl. Phys. **50**, 195103 (2017).

[11] D. Stange, N. von den Driesch, T. Zabel, F. Armand-Pilon, D. Rainko, B. Marzban, P. Zaumseil, J.-M. Hartmann, Z. Ikonic, G. Capellini, S. Mantl, H. Sigg, J. Witzens, D. Grützmacher, and D. Buca, ACS Photonics **5**, 4628 (2018).

[12] J. Chrétien, N. Pauc, F. Armand Pilon, M. Bertrand, Q.-M. Thai, L. Casiez, N. Bernier, H. Dansas, P. Gergaud, E. Delamadeleine, R. Khazaka, H. Sigg, J. Faist, A. Chelnokov, V. Reboud, J.-M. Hartmann, and V. Calvo, ACS Photonics **6**, 2462 (2019).

[13] Y. Zhou, Y. Miao, S. Ojo, H. Tran, G. Abernathy, J. M. Grant, S. Amoah, G. Salamo, W. Du, J. Liu, J. Margetis, J. Tolle, Y.-H. Zhang, G. Sun, R. A. Soref, B. Li, and S.-Q. Yu, Optica **7**, 924 (2020).

[14] N. Von Den Driesch, D. Stange, S. Wirths, G. Mussler, B. Holländer, Z. Ikonic, J. M. Hartmann, T. Stoica, S. Mantl, D. Grützmacher, and D. Buca, Chem. Mater. **27**, 4693 (2015).

[15] R. W. Millar, D. C. S. Dumas, K. F. Gallacher, P. Jahandar, C. MacGregor, M. Myronov, and D. J. Paul, Opt. Express **25**, 25374 (2017).

[16] D. Stange, S. Wirths, R. Geiger, C. Schulte-Braucks, B. Marzban, N. V. Den Driesch, G. Mussler, T. Zabel, T. Stoica, J.-M. Hartmann, S. Mantl, Z. Ikonic, D. Grützmacher, H.

Sigg, J. Witzens, and D. Buca, ACS Photonics **3**, 1279 (2016).

[17] M. R. M. Atalla, S. Assali, A. Attiaoui, C. Lemieux-Leduc, A. Kumar, S. Abdi, and O. Moutanabbir, (2020).

[18] É. Bouthillier, S. Assali, J. Nicolas, and O. Moutanabbir, Semicond. Sci. Technol. **35**, 095006 (2020).

[19] S. Assali, M. Elsayed, J. Nicolas, M. O. Liedke, A. Wagner, M. Butterling, R. Krause-Rehberg, and O. Moutanabbir, Appl. Phys. Lett. **114**, 251907 (2019).

[20] T. B. Bahder, Phys. Rev. B **46**, 9913 (1992).

[21] G.-E. Chang, S.-W. Chang, and S. L. Chuang, IEEE J. Quantum Electron. **46**, 1813 (2010).

[22] G. L. Bir and G. E. Pikus, *Symmetry and Strain- Induced Effects in Semiconductors. Translated from Russian by P. Shelnitz* (Wiley, New York, 1974).

[23] C. G. Van de Walle and R. M. Martin, Phys. Rev. B **34**, 5621 (1986).

[24] C. G. de Walle, Phys. Rev. B **39**, 1871 (1989).

[25] A. Gassenq, L. Milord, J. Aubin, K. Guilloy, S. Tardif, N. Pauc, J. Rothman, A. Chelnokov, J. M. Hartmann, V. Reboud, and V. Calvo, Appl. Phys. Lett. **109**, 242107 (2016).

[26] H. Lin, R. Chen, W. Lu, Y. Huo, T. I. Kamins, and J. S. Harris, Appl. Phys. Lett. **100**, 102109 (2012).

[27] K. Lu Low, Y. Yang, G. Han, W. Fan, and Y.-C. Yeo, J. Appl. Phys. **112**, 103715 (2012).

[28] H. Tran, W. Du, S. A. Ghetmiri, A. Mosleh, G. Sun, R. A. Soref, J. Margetis, J. Tolle, B. Li, H. A. Naseem, and S.-Q. Yu, J. Appl. Phys. **119**, 103106 (2016).

[29] T. Schmidt, K. Lischka, and W. Zulehner, Phys. Rev. B **45**, 8989 (1992).

[30] U. Kaufmann, M. Kunzer, M. Maier, H. Obloh, A. Ramakrishnan, B. Santic, and P. Schlotter, Appl. Phys. Lett. **72**, 1326 (1998).

[31] S. Assali, J. Greil, I. Zardo, A. Belabbes, M. W. A. De Moor, S. Koelling, P. M. Koenraad, F. Bechstedt, E. P. A. M. Bakkers, and J. E. M. Haverkort, J. Appl. Phys. **120**, (2016).

[32] E. Burstein, Phys. Rev. **93**, 632 (1954).

[33] E. F. Schubert, E. O. Göbel, Y. Horikoshi, K. Ploog, and H. J. Queisser, Phys. Rev. B **30**, 813 (1984).

[34] D. Ouadjaout and Y. Marfaing, Phys. Rev. B **41**, 12096 (1990).

[35] L. Viña, S. Logothetidis, and M. Cardona, Phys. Rev. B **30**, 1979 (1984).

[36] P. Y. Varshni, Physica **34**, 149 (1967).

[37] O. Madelung, U. Rössler, and M. Schulz, editors, *Group IV Elements, IV-IV and III-V Compounds. Part b - Electronic, Transport, Optical and Other Properties* (Springer-

Verlag, Berlin/Heidelberg, 2002).

[38] F. Pezzoli, A. Giorgioni, D. Patchett, and M. Myronov, ACS Photonics **3**, 2004 (2016).

[39] M. A. Reshchikov, J. Appl. Phys. **115**, 012010 (2014).

[40] S. Assali, A. Dijkstra, A. Li, S. Koelling, M. A. Verheijen, L. Gagliano, N. von den Driesch, D. Buca, P. M. Koenraad, J. E. M. Haverkort, and E. P. A. M. Bakkers, Nano Lett. **17**, 1538 (2017).

[41] M. Leroux, N. Grandjean, B. Beaumont, G. Nataf, F. Semond, J. Massies, and P. Gibart, J. Appl. Phys. **86**, 3721 (1999).

[42] S. Koelling, A. Li, A. Cavalli, S. Assali, D. Car, S. Gazibegovic, E. P. A. M. Bakkers, and P. M. Koenraad, Nano Lett. **17**, 599 (2017).

[43] W. P. Bai, N. Lu, A. Ritenour, M. L. Lee, D. A. Antoniadis, and D.-L. Kwong, IEEE Electron Device Lett. **27**, 175 (2006).

[44] M. Christian Petersen, A. Nylandsted Larsen, and A. Mesli, Phys. Rev. B **82**, 075203 (2010).

[45] C. G. Van de Walle and J. Neugebauer, J. Appl. Phys. **95**, 3851 (2004).

[46] D. Stange, S. Wirths, N. Von Den Driesch, G. Mussler, T. Stoica, Z. Ikonic, J. M. Hartmann, S. Mantl, D. Grützmacher, and D. Buca, ACS Photonics **2**, 1539 (2015).

[47] M. S. Seifner, A. Dijkstra, J. Bernardi, A. Steiger-Thirsfeld, M. Sistani, A. Lugstein, J. E. M. Haverkort, and S. Barth, ACS Nano acsnano.9b02843 (2019).

[48] J. Nicolas, S. Assali, S. Mukherjee, A. Lotnyk, and O. Moutanabbir, Cryst. Growth Des. acs. cgd.0c00270 (2020).

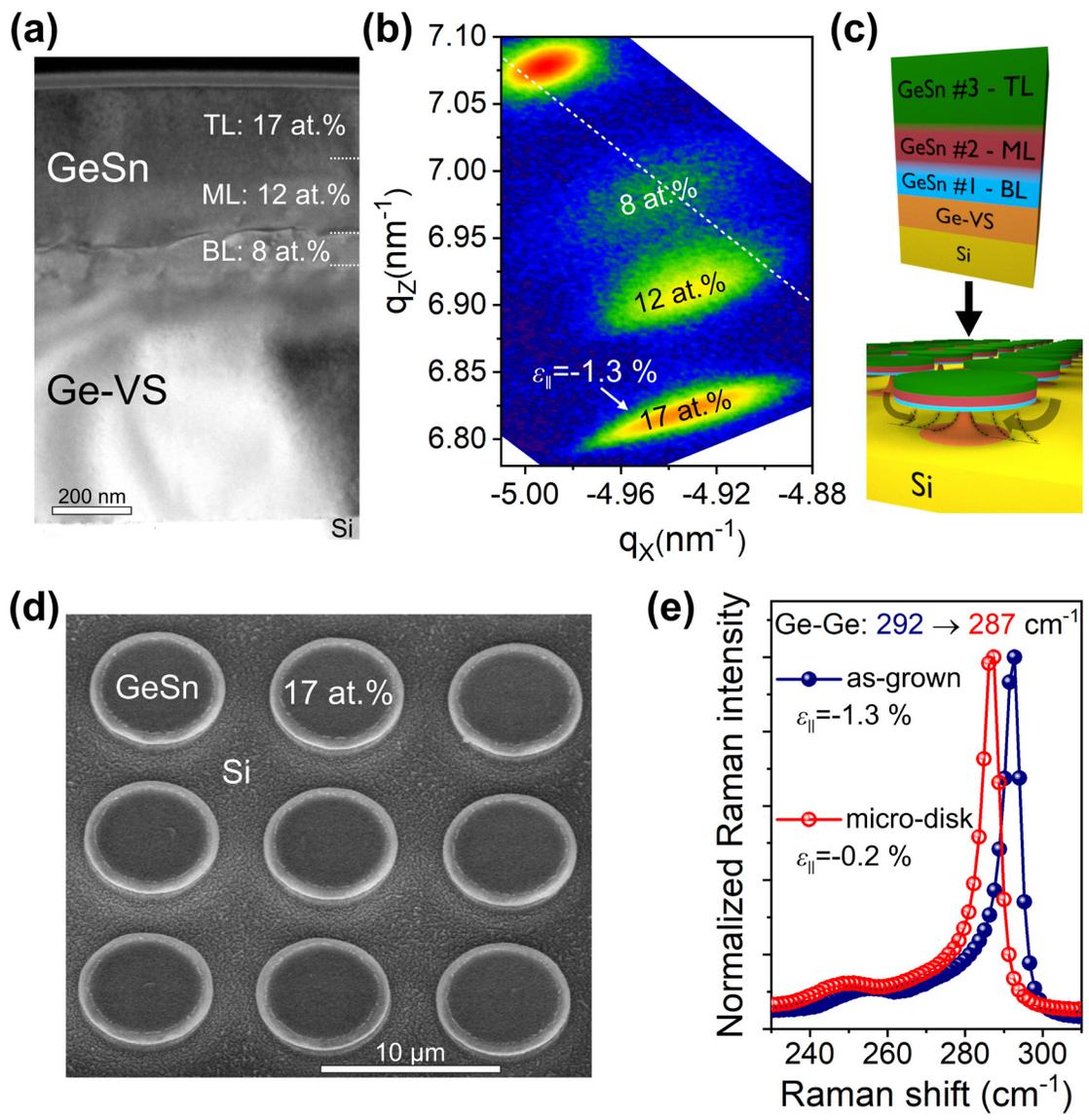

Figure 1

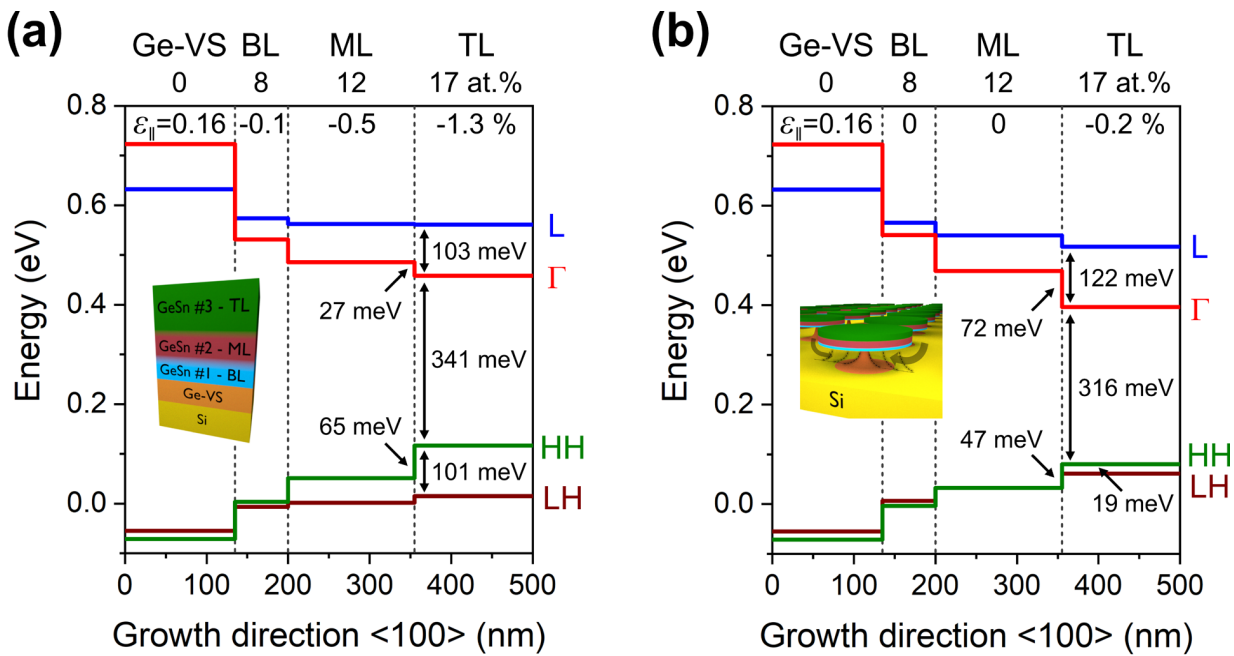

**Figure 2**

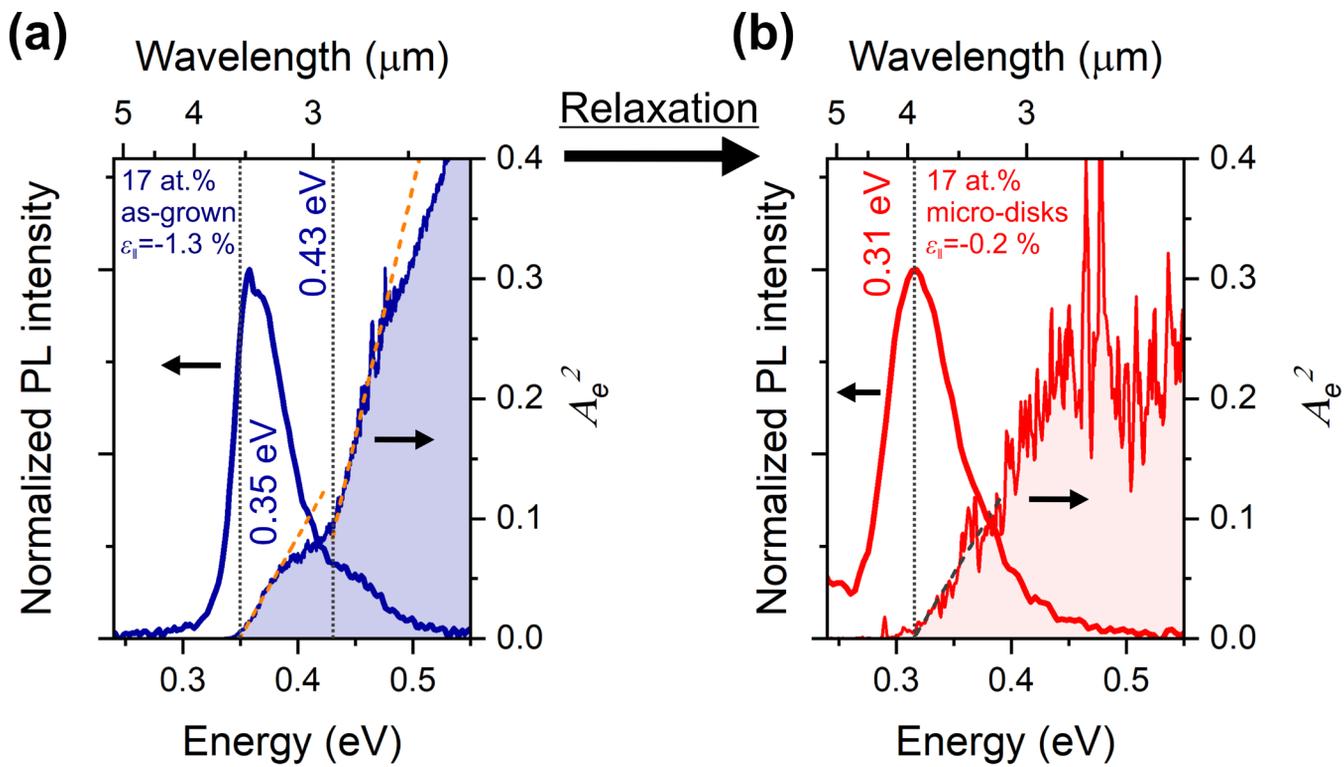

Figure 3

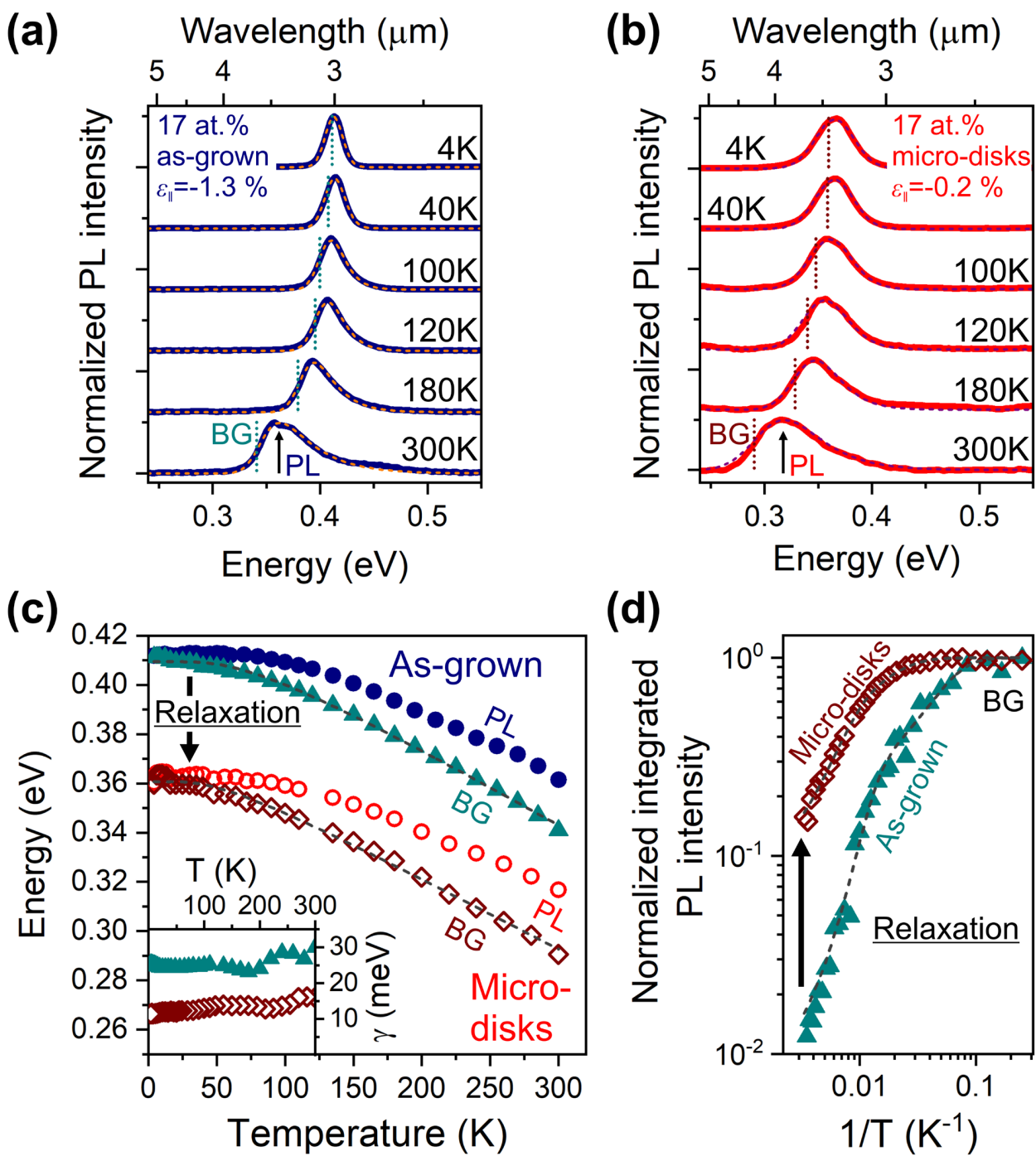

Figure 4

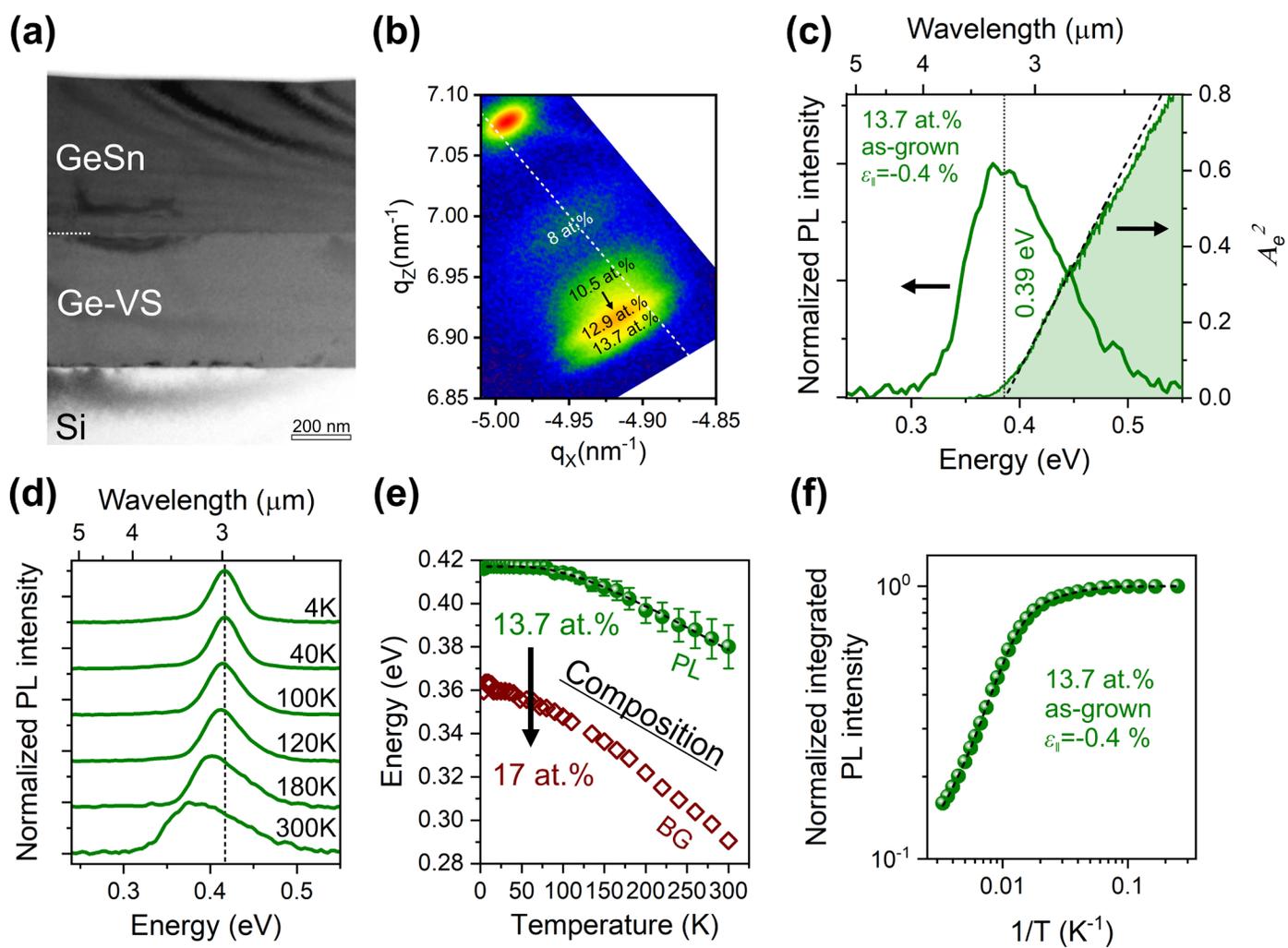

**Figure 5**

|  | $a$ (eV) | $b$ (eV) | $\theta_D$ (K) |
|---|---|---|---|
| As-grown 17.0 at.% | 0.443 ±0.002 | 0.034 ±0.002 | 209 ±10 |
| Micro-disks 17.0 at.% | 0.382 ±0.002 | 0.021 ±0.002 | 146 ±14 |
| As-grown 13.7 at.% | 0.455 ±0.004 | 0.040 ±0.004 | 436 ±24 |

**Table 1**

|  | $E_a^1$ (meV) | $c_1$ | $E_a^2$ (meV) | $c_2$ |
|---|---|---|---|---|
| As-grown 17 at.% | 3.8 ±0.5 | 4.2 ±0.8 | 32 ±7 | 199 ±160 |
| Micro-disks 17 at.% | 9 ±1 | 1.4 ±0.3 | 34 ±3 | 15 ±2 |
| As-grown 13.7 at.% | 4.7 ±0.2 | 0.42 ±0.02 | 25.6 ±0.2 | 13.5 ±0.1 |

**Table 2**

# Supplemental Material:

# Mid-infrared emission and absorption in strained and relaxed direct band gap GeSn semiconductors

S. Assali,[1,+,*] A. Dijkstra,[2,+] A. Attiaoui,[1] É. Bouthillier,[1] J. E. M. Haverkort,[2] and O. Moutanabbir[1]

[1] Department of Engineering Physics, École Polytechnique de Montréal, C. P. 6079, Succ. Centre-Ville, Montréal, Québec H3C 3A7, Canada

[2] Department of Applied Physics, Eindhoven University of Technology, 5600 MB Eindhoven, The Netherlands

## Contents

# S1. Description of the absorption setup.

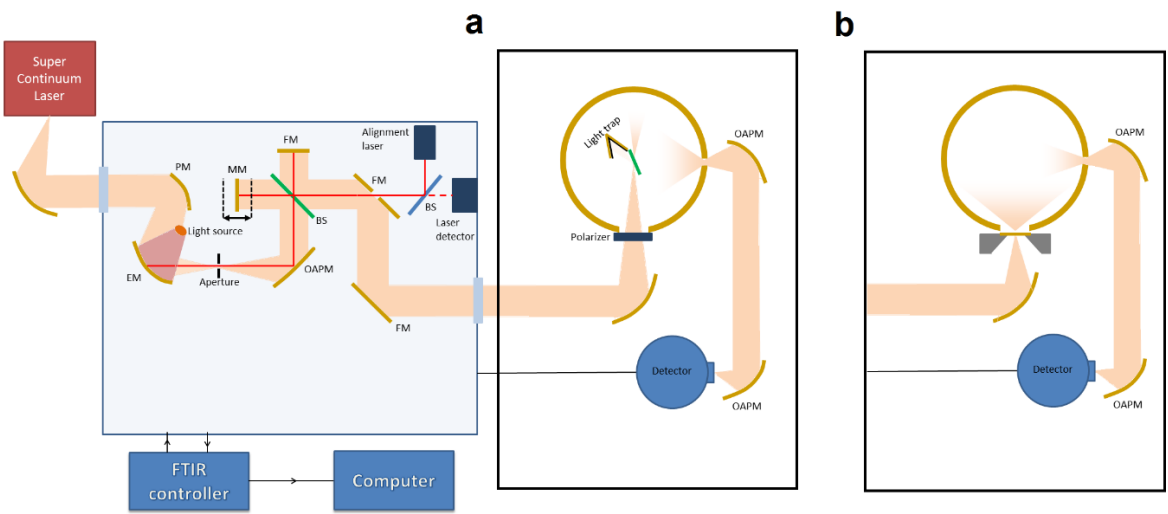

**Figure S1.** Schematics of the optical setup used for absorption measurements. A supercontinuum laser has been used as a white light source with emission up to 4.1 μm, or has been replaced by the internal glowbar of the FTIR system for measurements further in the infrared. The emission of the source was then coupled through the interferometer and shone onto the sample in two different geometries. In (**a**) the light is linearly polarized and focused onto the sample which is placed on a rotatable sample holder at the center of a gold-coated integrating sphere (IS). The light that is reflected from the front side of the sample was rejected by collecting it in a small light trap right next to the sample. The transmitted light is collected from the integrating sphere through a baffled port and focused onto an HgCdTe-detector. Alternatively in (**b**) the sample is mounted at the entrance of the integrating sphere and the light is focused normal to the sample surface. The transmitted light is then collected by the IS. Background measurements have been performed directly before each transmission measurement and the total transmittance was determined as $T_{tot} = I_{meas}/I_{background}$. The further analysis of the data is described in Supplemental Material S4.

## S2. Structural characterization of the Ge$_{0.83}$Sn$_{0.17}$ micro-disks.

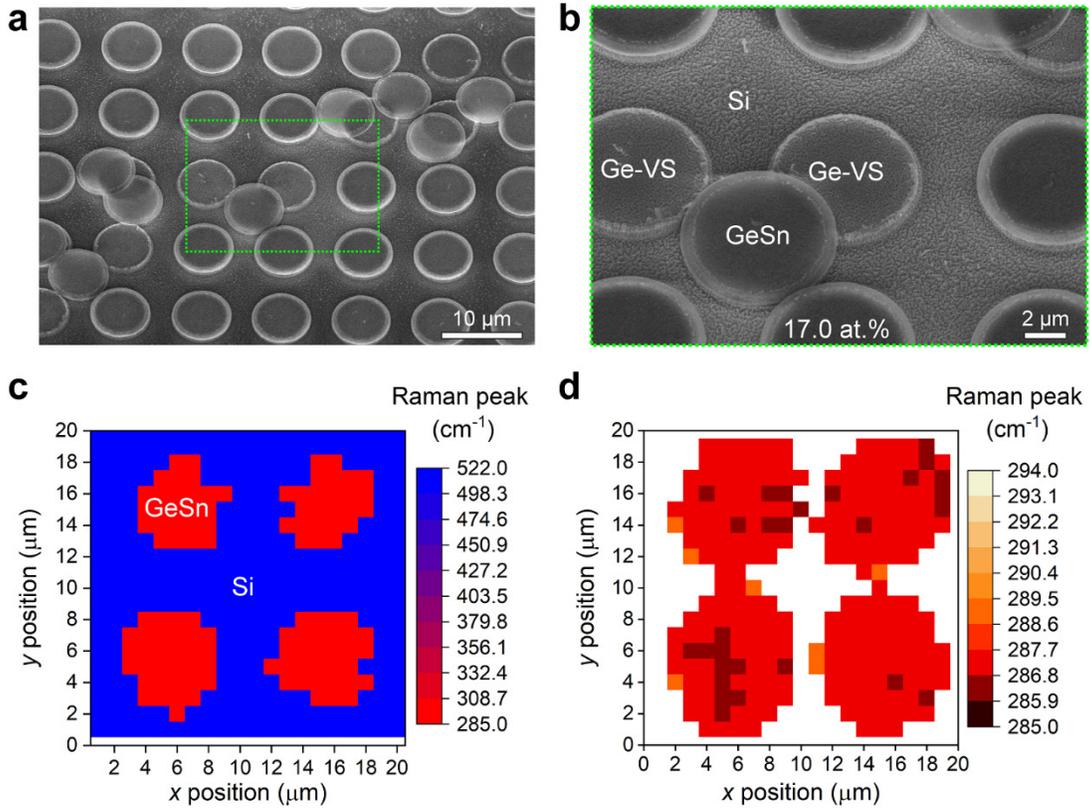

**Figure S2.** (**a-b**) SEM images the Ge$_{0.83}$Sn$_{0.17}$ micro-disk arrays (45° tilting angle) showing few micro-disks that are detached from the Ge-VS and redeposited in a different location. (**c-d**) Raman maps indicating the complete GeSn and Ge-VS removal in between the micro-disks (**c**) and the peak position across the individual micro-disks (**d**).

The filling factor (FF) of micro-disks is defined as the ratio of the relative area covered by the micro-disks:

$$FF = \frac{\pi d^2}{4p^2} \qquad \text{(eq. S1)}$$

Where *d* and *p* are the diameter and pitch of the micro-disks, respectively. From Fig. S2a, a FF~39 % is estimated.

## S3. 8-band *k·p* estimation of temperature-dependent bowing parameter

The temperature dependence of the GeSn band gap is a function of both composition and strain in the material. However, in the available literature on this topic multiple assumptions have been made, which provide a rather simplified picture. For instance, in the work of L. Qian et al. [1] the Varshni coefficient of Ge was used for $Ge_{0.88}Sn_{0.12}$. Moreover, in the work of J. Hart et al. [2] the authors developed a Varshni dependence of the band gap as a function of the Sn content, which led to an estimated 300 K direct band gap of 0.529 eV in $Ge_{0.89}Sn_{0.11}$ with a compressive strain of -1.4 %. The strain variation with decreasing temperature due to thermal expansion mismatch was neglected in the previous work.

In our *k·p* method a strain-independent bowing parameter is selected (2.42 eV), resulting in a band gap of 0.492 eV at 300 K in a compressively strained $Ge_{0.99}Sn_{0.11}$ layer. This translates into a relative error of 7.4 % between the calculated and measured value. At 20 K, using Varshni law, the measured band gap for the same composition was estimated to be 0.594 eV, while our 8-band *k·p* method predicts a band gap of 0.572 eV, with a relative error of 3.8 %, where the same deformation potential constant as in the work of H. Lin et al. [3] was selected. Thus, using a fixed bowing parameter $b_\Gamma$=2.5 eV and $b_L$=0.8 eV, the measured band gaps are always underestimated by a few percent. To fix this issue, a simple fitting routine to estimate the bowing parameter defined in Eq. S2 was used, where the measured PL direct band gap was fitted with the 8-band *k.p* at each measured temperature which allowed for a point-by-point estimation of the bowing parameter at each Sn content and strain. This allows for a reduction of the relative error (below 1%) between the measured and the simulated band gaps.

The compositional dependence of the $Ge_{1-y}Sn_y$ bandgaps was calculated using a standard quadratic equation of the form:

$$E_i^{Ge_{1-y}Sn_y}(y,T) = yE_i^{Sn}(y,T) + (1-y)E_i^{Ge}(y,T) - y(1-y)b_i(\varepsilon,T) \quad \text{(eq. S2)}$$

where $i = \Gamma, L$ for the direct and indirect bandgaps, respectively, with the $E_i$ values for pure Ge and α-Sn given in Ref. [4] and Ref. [5], respectively.

The use of the band gap bowing parameter available in literature induces the high relative error between the measured and calculated band gap for the Ge$_{0.83}$Sn$_{0.17}$. As this bowing parameter depends on strain, composition, and temperature, thus finding a universal expression of the bowing parameter is out of the scope of this work. However, it is possible to extract a new expression of the bowing parameters for the strained and relaxed Ge$_{0.83}$Sn$_{0.17}$. To that end, the measured PL band gap of the micro-disk, the as grown Ge$_{0.83}$Sn$_{0.17}$ and the relaxed Ge$_{0.863}$Sn$_{0.137}$ were used to fit the calculated $k \cdot p$ direct band gap in order to extract the bowing parameter $b_\Gamma(T)$. The bowing parameters were then fitted to a second order polynomial function:

$$b_\Gamma^{relaxed-17\%Sn}(T) = 1.85 + 8.32 \times 10^{-4}\,T + 3.02 \times 10^{-6}\,T^2 \ (eV) \quad \text{(eq. S3)}$$

$$b_\Gamma^{strained-17\%Sn}(T) = 2.16 + 6.24 \times 10^{-4}\,T + 3.50 \times 10^{-6}\,T^2 \ (eV) \quad \text{(eq. S4)}$$

$$b_\Gamma^{relaxed-13.7\%Sn}(T) = 2.40 + 9.43 \times 10^{-4}\,T + 2.47 \times 10^{-6}\,T^2 \ (eV) \quad \text{(eq. S5)}$$

These new band gap bowing parameters will give a relative error of less than 1 % between the measured PL and the simulated direct band gap. The fitted direct band gaps for the strained and relaxed Ge$_{0.83}$Sn$_{0.17}$ are shown in Figure S3a together with the measured PL maxima. It is important to mention that the deformation potential constants (DPC) of GeSn were linearly interpolated between those of Ge [4] and $\alpha$-Sn [5]. Furthermore, the DPCs are assumed to be temperature independent.

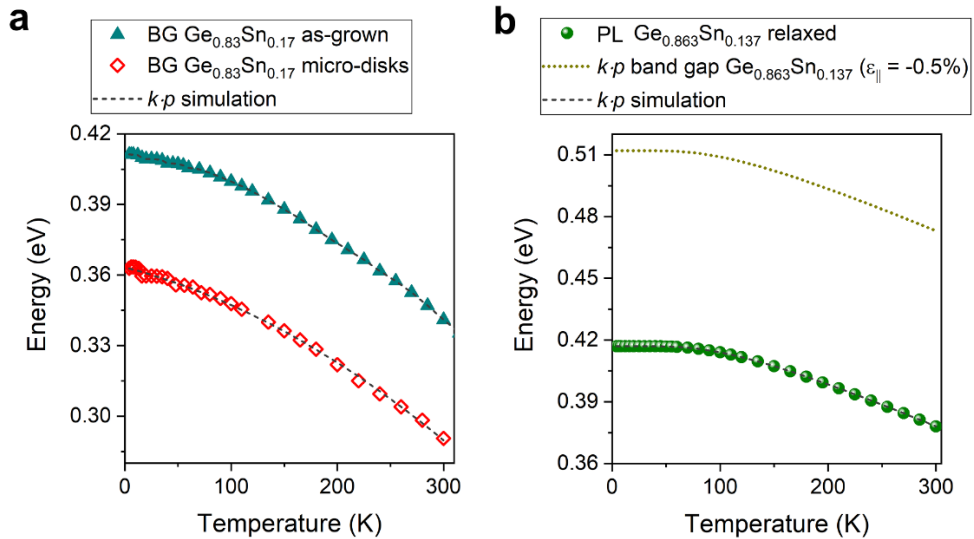

**Figure S3.** (**a-b**) Temperature-dependent direct band gap for (**a**) as-grown and relaxed Ge$_{0.83}$Sn$_{0.17}$ estimated using the spectral line-shape model for PL spectra fit and (**b**) the Ge$_{0.863}$Sn$_{0.137}$ layer estimated using the maxima in the PL spectra. The 8-band $k \cdot p$ simulations are shown using dashed curves.

## S4. Determination of the absorbance from transmission measurements

To interpret how the absorption coefficients, follow from the measured transmission $T_{tot}$, the propagation of a ray of light through a sample is considered as shown in Fig. S4. The two considered processes are, absorption when a ray of light travels through a layer resulting in transmission $t_i$ and secondly, reflection at every refractive-index-changing interface $R_{i,j}$.

We will first consider measurements performed on as-grown samples as depicted in Fig. S4a. To avoid interference fringes all measurements on these samples have been performed at the Brewster angle with p-polarized light which means that the reflection from the sample surface is theoretically zero.

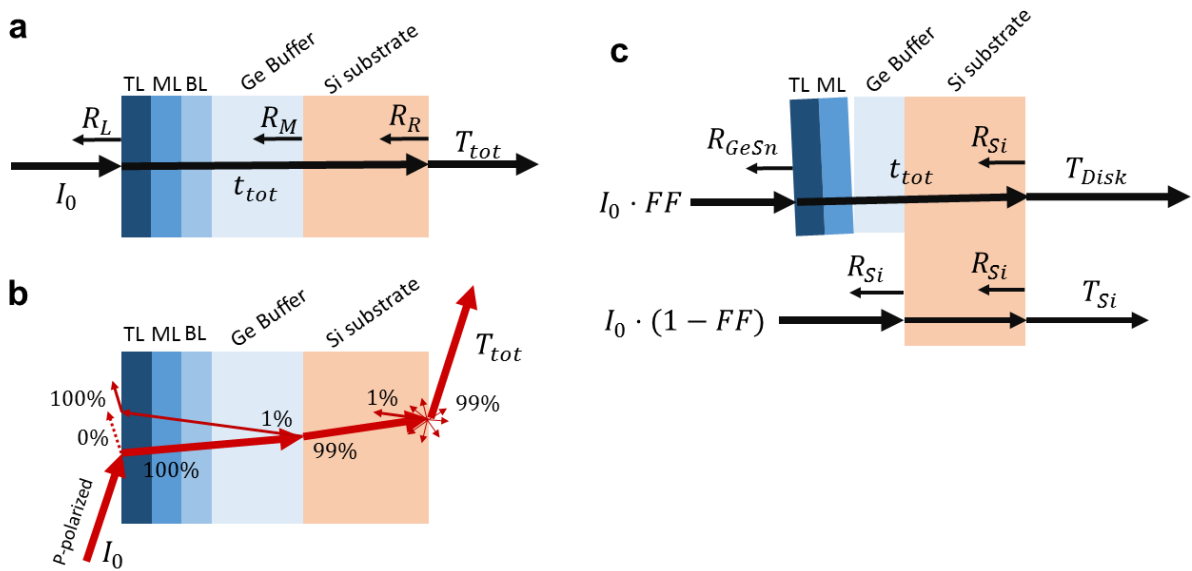

**Figure S4.** Schematics of all reflections taken into account for the determination of the total transmission $T_{tot}$. Where in (**a**) the considered reflections are schematically shown for the as-grown sample. In (**b**) the same situation is considered as in (**a**) but now for the Brewster angle specifically, where almost all light directly passes through the whole sample and multiple reflections can be neglected. In (**c**) the situation is shown for the micro-disk sample where only a fraction of the light, equal to the fill factor (FF), interacts with the disks and 1-FF bypasses the disks and interacts with the silicon substrate. Here it is assumed that all interfaces are smooth and that the disks have fully collapsed onto the layer underneath such that the Fresnel equations are valid for the calculation of reflections.

Despite this Brewster condition a small reflection from the sample surface is still observed (indicated with a dashed arrow in Fig. S4b) which could be caused by small amounts of Sn segregation at the surface of the sample, a slight surface roughness or the finite angular distribution of the incoming light. Because GeSn and Ge have an almost identical refractive index of ~4.2 and ~4 respectively, reflections between the GeSn-layers and the Ge buffer layer are negligible. Using the Fresnel equations for p-polarized light, a mere 1% reflection is expected at the Ge to Si interface (Fig. S4b). This 1% reflection however is perfectly transmitted at the sample surface when travelling to the left due to the Brewster condition and can therefore interfere with the reflection from the sample surface. To avoid these interference effects all light leaving through the front of the sample is rejected from the integrating sphere by using a light trap as schematically illustrated in Fig. S1a. The successful suppression of interference fringes is demonstrated in Fig. S6. The light that travels to the right through the silicon layer is expected to be transmitted out of the sample by 99 % because the Brewster angles for GeSn (n=4.2, $\theta_{brewster}$=76.6º) and Si (n=3.4, $\theta_{brewster}$=73.6º) are very similar. However, the real transmittance will be reduced due to scattering on the unpolished backside of the sample. Despite this limitation one can assume that all light is either rejected on the front side of the sample or travels through the whole sample just once.

The transmission through a layer $i$ is given by $t_i = \exp(-\alpha_i \cdot l_i)$ where $\alpha_i$ is the absorption coefficient of the respective layer and $l_i$ is the distance that the light travels through the layer given by $l_i = d_i/\cos(\theta_i)$, with $d_i$ the thickness of the layer and $\theta_i$ the angle with respect to the surface normal in which the light travels. Because the refractive indices of the GeSn-layers and Ge-layer are almost identical, multiple reflections between these layers can be neglected. This claim is supported by the fringe spacing shown for off-Brewster angle measurements in Fig S6d-f which match the thickness of the complete GeSn stack plus the Ge buffer layer. Therefore, light travelling through the stack will have a net transmission that is a simple product of the transmissions of the different layers, given by equation S6.

$$t_{tot} = \prod_i t_i = \exp\left(\sum_i -\alpha_i \cdot l_i\right) = \exp(-A_e) \qquad \text{(eq. S6)}$$

In equation S6 also the Napierian absorbance $A_e$ is introduced, defined as in equation S7. The Napierian absorbance is a variable that linearly scales with the absorption coefficient $\alpha_i$ but is experimentally more easily accessible yet can be interpreted similarly.

$$A_e = \sum_i A_i = \sum_i \alpha_i \cdot l_i \qquad \text{(eq. S7)}$$

As a last step the Napierian absorbance needs to be calculated from the measured transmittance, as plotted in Fig. S5a-c. The process is schematically depicted in Fig. S4b, because of the Brewster angle, in principle, no reflections are expected which suggests that the total transmittance is given by $T_{tot} = t_{tot}$. However, due to a small surface roughness on the front side of the sample a small reflection from the front side of the sample is lost in the light trap. Additionally, the unpolished backside of the sample will scatter the light, of which a fraction will also disappear in the light trap. These two scattering losses can be summarized in a $I_o \cdot b$ term. Therefore, the total transmittance is given by equation S8.

$$T_{tot} = (1 - b) \cdot \exp(-A_e) \qquad \text{(eq. S8)}$$

From which the Napierian absorbance can be determined to be given by equation S9, in which $b' = -\ln(1-b)$ is a baseline that needs to be subtracted to find the actual $A_e$ spectrum.

$$A_e = -\ln(T_{tot}) - b' \qquad \text{(eq. S9)}$$

The baseline correction $b'$ is performed by averaging the values of the absorption coefficient at the low energy side of the spectrum, below the band gap of the top layer, and subtracting this value from the full spectrum.

Now we shift to the micro-disks, which have been measured in a normal-incidence experiment (schematically depicted in Fig S4c) therefore additional reflections need to be taken into account with respect to the as-grown samples. From Fig S6b and S6e it follows that no interference fringes are observed independent of the angle of incidence. The reflections $R_{i,j}$ at the interfaces are determined using the Fresnel equations for perpendicular incidence given in equation S10 when light travels from medium $i$ to $j$. Here $n_i$ and $n_j$ are the refractive indices of the respective media.

$$R_{i,j} = \left| \frac{n_i - n_j}{n_i + n_j} \right|^2 \qquad \text{(eq. S10)}$$

In the case of a planar sample, when taking into account multiple reflections on both sides of the full sample, but neglecting reflections in between the layers inside the sample, the general expression S11 is found [6,7] for the total transmittance $T$, which is the net transmittance through a layer stack in which the refractive index changes gradually.

$$T = \frac{(1 - R_L) \cdot (1 - R_R) \cdot t_{tot}}{1 - t_{tot}^2 \cdot R_L \cdot R_R} \qquad \text{(eq. S11)}$$

In which $R_L$ is the reflection on the left side of the sample and $R_R$ the reflection on the right side. To apply this expression to the micro-disk sample we note that the fraction of light interacting with the micro-disks is given by the fill factor (FF), which means that an intensity of $I_0 \cdot FF$ falls onto the disks and an intensity of $I_0 \cdot (1 - FF)$ falls past the disks as also depicted in Fig S4c. The net transmittance through the micro-disk sample is then given by equation S12:

$$T_{tot} = T_{Disk} + T_{Si} = \frac{FF \cdot (1 - R_L) \cdot (1 - R_R) \cdot t_{tot}}{1 - t_{tot}^2 \cdot R_L \cdot R_R} + \frac{(1 - FF) \cdot (1 - R_{Si})^2}{1 - R_{Si}^2} \qquad \text{(eq. S12)}$$

Here it is assumed that the micro-disks have fully collapsed and are lying on a thin germanium layer as shown in Fig S3c, that there is no significant absorption in the silicon layer (i.e. $t_{si} = 1$) and that scattering effects due to surface roughness can be neglected. Equation S12 can be rewritten to explicitly express $t_{tot}$ in terms of the measured $T_{tot}$ and reflections that can be calculated from refractive indices using equation S6. The Napierian absorbance is then calculated using equation S6. The final result still required a small baseline correction which we attribute due to scattering at rough surfaces which cannot be taken into account with this analysis. The correction is performed analogous with equation S9. We note that the baseline corrections shows that there is a small error in the absolute value of the absorbance, however the main purpose is to determine the band gaps which is only influenced by the shape of the absorbance spectrum.

# S5. Transmittance and absorbance spectra.

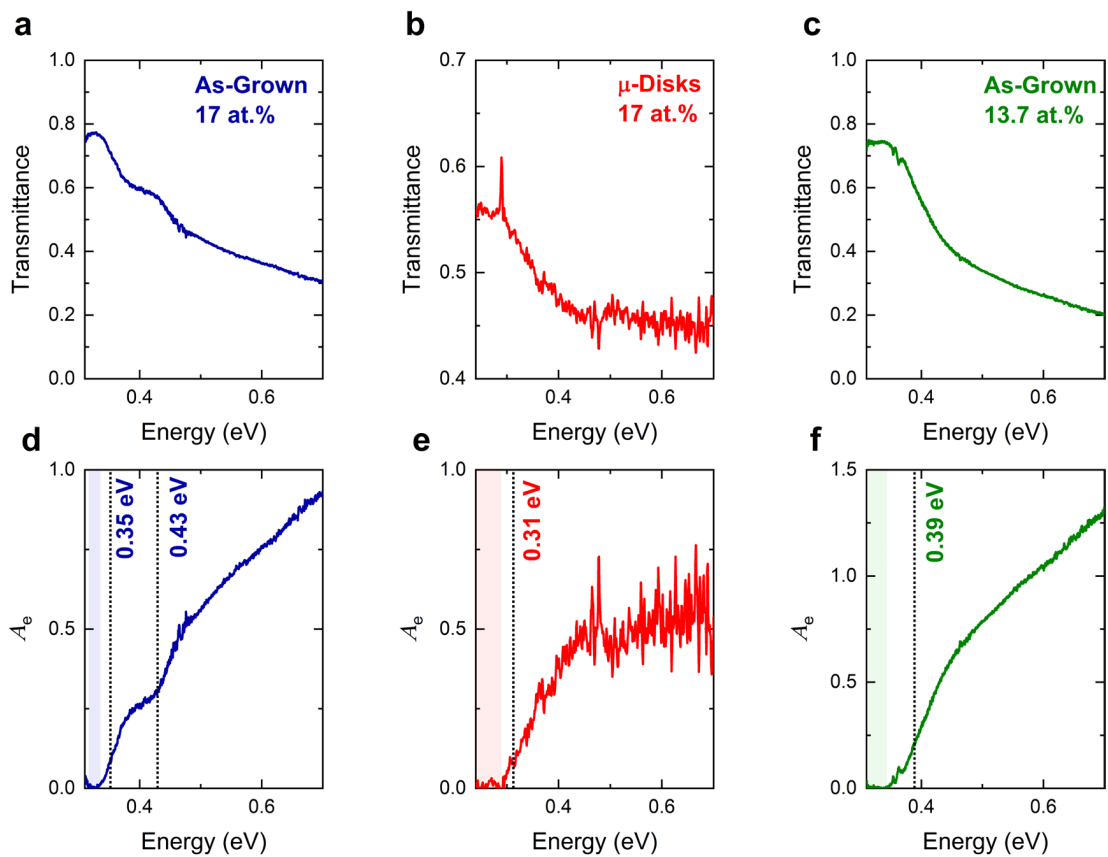

**Figure S5.** In (**a**), (**b**) and (**c**) the measured transmittance spectra of respectively the as-grown $Ge_{0.83}Sn_{0.17}$ sample, the $Ge_{0.83}Sn_{0.17}$ micro-disks and the as-grown $Ge_{0.863}Sn_{0.137}$ samples are shown. Where the spectra shown in (**a**) and (**c**) have been measured using a supercontinuum source at the Brewster angle as shown in Fig. S1a and the data in (**b**) were collected using a glow bar at normal incidence, schematically shown in Fig. S1b. In (**d**), (**e**) and (**f**) the Napierian absorbance as calculated from the data in figures (**a**), (**b**) and (**c**) is shown using formula S9 for (**d**) and (**f**) and equation S12 for (**e**). Baseline corrections have been performed by averaging the low energy part of the spectrum (marked as a shaded region) and subtracting this value. Due to the small uncertainty in determining the baseline the absorbance is subject to an absolute error of <0.1.

## S6. Suppression of interference fringes in absorbance spectra.

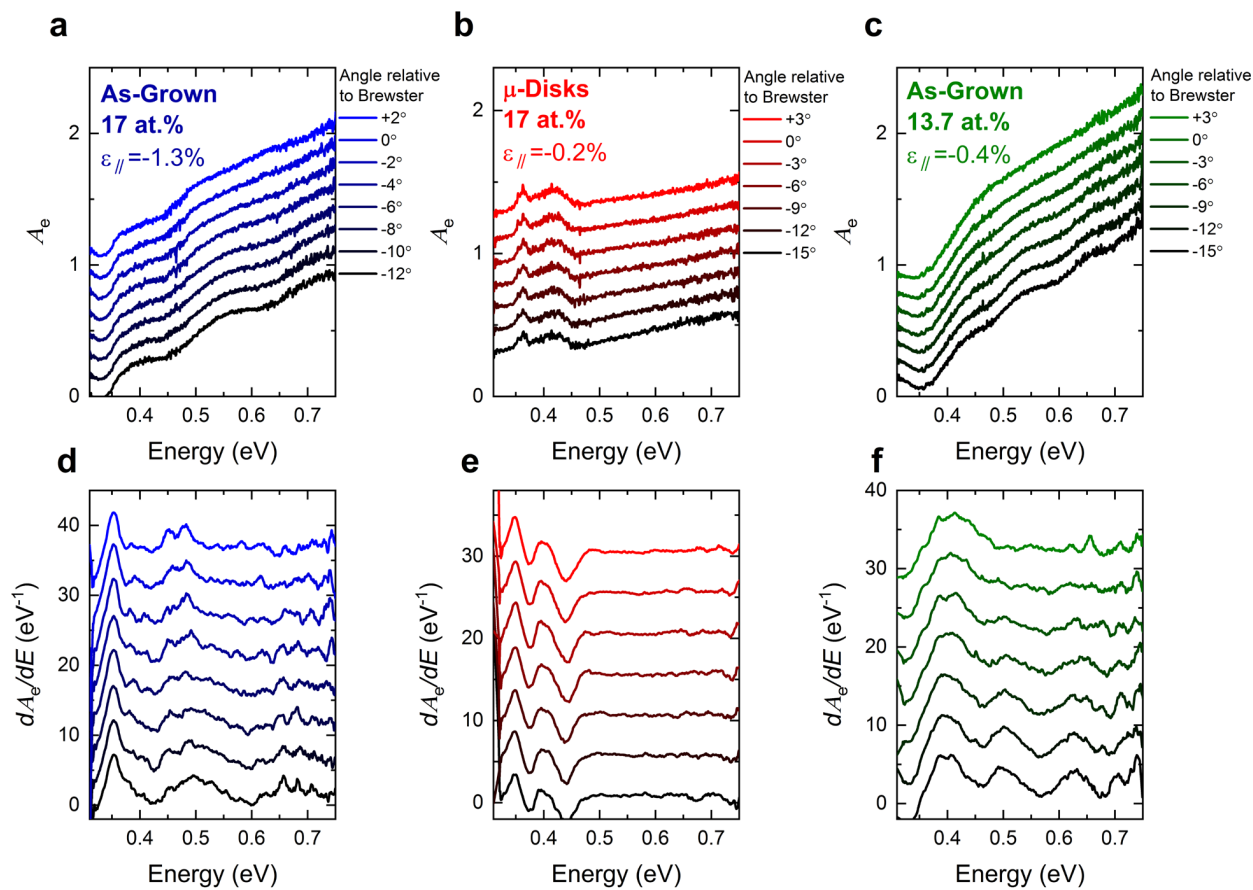

**Figure S6.** (**a**), (**b**) and (**c**) show the Napierian absorbance at varying angles relative to the Brewster angle of respectively the as-grown $Ge_{0.83}Sn_{0.17}$ sample, the $Ge_{0.83}Sn_{0.17}$ micro-disks and the as-grown $Ge_{0.863}Sn_{0.137}$ samples, measured using a supercontinuum (SC) light source in the configuration as shown in Fig. S1. In (**a**) and (**c**), clearly interference fringes emerge when the sample is rotated away from the Brewster angle. In both cases the fringe spacing matches the thickness of the GeSn layers and Ge buffer layer combined. This also demonstrates that the onset of the band gap in (**a**) and (**c**) at 0.35 eV and 0.39 eV respectively are real and not a consequence of any fringes. In (**b**) (data corresponding to the $Ge_{0.83}Sn_{0.17}$ micro-disks), interference fringes remain absent independent of the angle, probably due to additional scattering. No band gap is identified because it lies outside of the spectral window of the here used SC light source. The small feature around 0.36 eV is due to an absorption line in air. To highlight the features of all spectra the derivatives of the Napierian absorbance curves are plotted in (**d**), (**e**) and (**f**) corresponding to the data in (**a**), (**b**) and (**c**) respectively.

## S7. Power-dependent PL measurements on as-grown Ge$_{0.83}$Sn$_{0.17}$ at 4K.

The power-dependent PL spectra acquired in the 6.9 W/cm$^2$ to 5.4 kW/cm$^2$ range are plotted in Fig. S7a, while the integrated PL intensity ($I_{PL}$) and emission energy are plotted as function of the excitation power density ($P_{EXC}$) in Fig. S7b, top panel. Free- and bound-exciton recombination [8–10] is observed at low power with a slope $m\sim 1$, which is extracted from fitting the data with the power law $I_{PL} \propto P_{EXC}^m$, and a constant emission energy. At higher excitation power band to band emission is visible, where a Burstein-Moss effect [11] with blue-shift of 2-6 meV/decade is estimated.

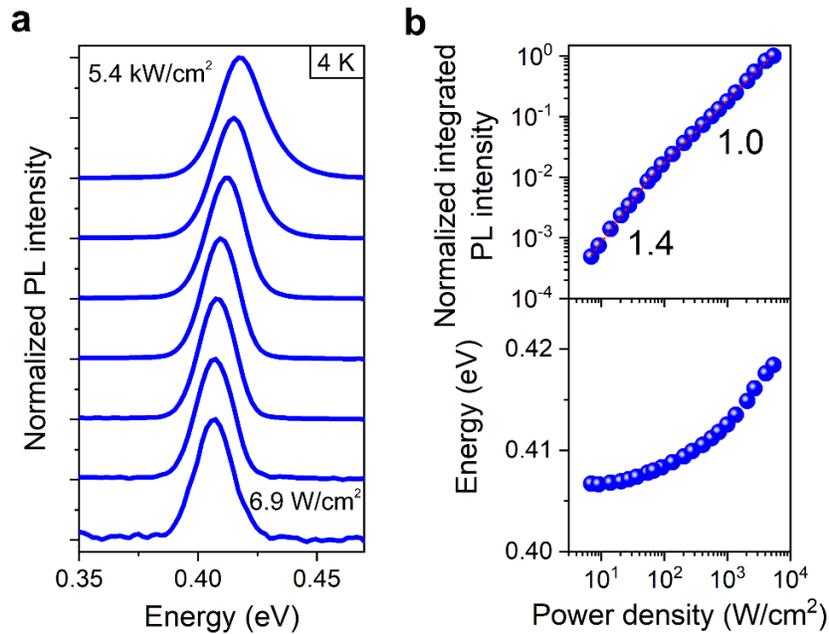

**Figure S7.** (**a**) PL spectra, normalized to their own maximum, acquired using excitation power densities from 6.9 W/cm$^2$ to 5.4 kW/cm$^2$. (**b-c**) Plot of the integrated PL intensity, normalized to its integrated intensity at the highest excitation density (b) and emission energy (c) as a function of the excitation power density.

## S8. Spectral line-shape model for PL spectra fit

To estimate the band gap from the photoluminescence measurements a basic spectral line-shape model is employed and fitted to the spectra. The used model is the joint density-of-states (JDOS) multiplied with a Boltzmann distribution, resulting in equation S13 which can also be found in several textbooks [12–14].

$$I_{PL} \propto \sqrt{E - E_g} \cdot \exp\left(-\frac{E}{k_B T}\right) \qquad \text{S13}$$

In which $E_g$ is the band gap energy, $T$ is the temperature and $k_B$ is the Boltzmann constant. Equation S13 is valid for intrinsic semiconductors at low excitation such that the Fermi-Dirac equation can be estimated with a Boltzmann distribution. Additionally, the usage of the expression for the JDOS $\propto \sqrt{E - E_g}$ assumes parabolic bands with a $\boldsymbol{k}$-independent matrix element. Examples of the spectrum based on equation S13 are shown in Fig. S7 as blue curves for different temperatures.

However, this very simple model assumes a hard onset of the JDOS where in reality imperfections in the crystal give rise to an Urbach tail that broadens the PL spectrum on the low energy side. Moreover, the measured spectra suffer from alloy-broadening which gives a gaussian [15,16] smearing of the net luminescence emission. To take these additional broadening mechanisms into account the simple line-shape model given by equation S13 is convoluted with a gaussian function resulting in the complete model used in this work given by equation S14.

$$I_{PL} = A \cdot \left[\sqrt{E - E_g} \cdot \exp\left(-\frac{E}{k_B T}\right)\right] * \left[\frac{1}{\gamma\sqrt{2\pi}} \exp\left(-\frac{E^2}{2\gamma^2}\right)\right] \qquad \text{S14}$$

In this relation $\gamma$ is a broadening parameter describing the width of the Gaussian and $A$ is a scaling constant. In Fig. S8 the effect of the convolution with a gaussian function (depicted in green) on the eventual PL spectrum (depicted in red) is illustrated at different temperatures. At the lowest temperature of 20K the shape of the spectrum is completely dominated by the gaussian part of this model. Only for elevated temperatures (typically higher than ~50 K) the PL spectrum asymmetrically broadens due to a Boltzmann tail on the high energy side. Therefore, the temperature cannot act as a reliable fitting parameter over the full temperature range. For this reason the temperature in this fitting model has been set equal to the lattice temperature of the sample and $E_g, A$ and $\gamma$ have been used as fitting parameters. Close agreement of this model with experiment can be clearly seen in Fig. 4a,b of the main text.

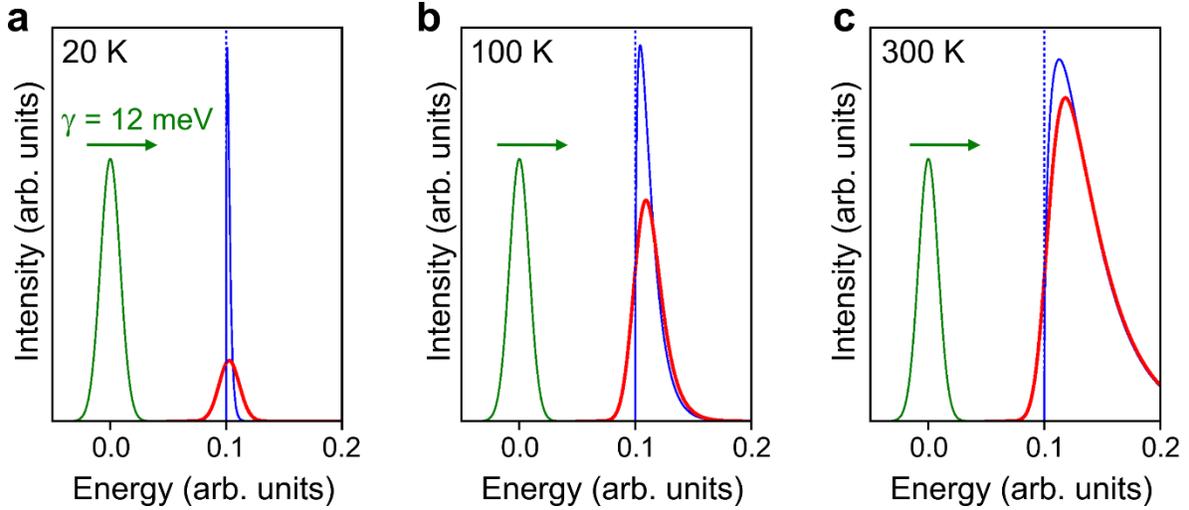

**Figure S8. (a-c)** As-grown Ge$_{0.83}$Sn$_{0.17}$ PL spectra acquired at 20 K (a), 100 K (b) and 300 K (c). All blue curves are based on equation S13 for which an example band gap of 0.1 eV is chosen (also indicated as a vertical dashed line) and an arbitrary scaling. The red curves depict examples of the full model given by equation S14, obtained by convoluting the model based on equation S13 (blue curves) with a gaussian (green curve). A value of $\gamma$ = 12 meV was chosen to closely resemble the experimental data.

## S9. Absence of impurities in atom probe tomography (APT) measurement.

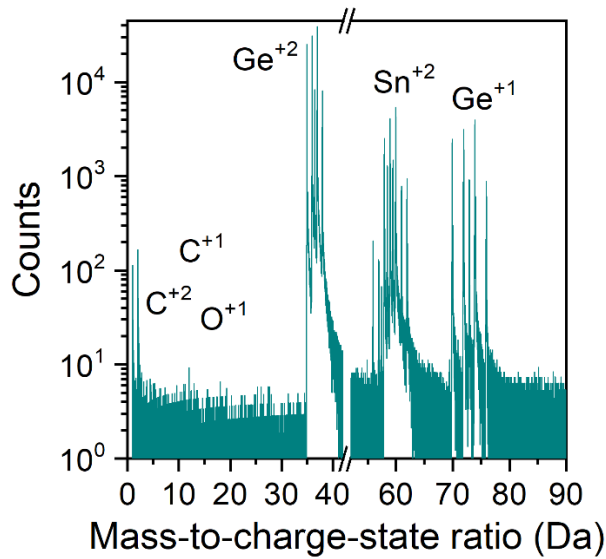

**Figure S9.** Mass spectrum extracted from the APT measurements of Ge$_{0.83}$Sn$_{0.17}$ from Ref. [17].

# References

[1] L. Qian, W. J. Fan, C. S. Tan, and D. H. Zhang, Opt. Mater. Express **7**, 800 (2017).

[2] J. Hart, T. Adam, Y. Kim, Y.-C. Huang, A. Reznicek, R. Hazbun, J. Gupta, and J. Kolodzey, J. Appl. Phys. **119**, 093105 (2016).

[3] H. Lin, R. Chen, W. Lu, Y. Huo, T. I. Kamins, and J. S. Harris, Appl. Phys. Lett. **100**, 102109 (2012).

[4] D. J. Paul, J. Appl. Phys. **120**, 043103 (2016).

[5] J. R. Chelikowsky and M. L. Cohen, Phys. Rev. B **14**, 556 (1976).

[6] V. Ariel, V. Garber, G. Bahir, S. Krishnamurthy, and A. Sher, Appl. Phys. Lett. **69**, 1864 (1996).

[7] E. Finkman and S. E. Schacham, J. Appl. Phys. **56**, 2896 (1984).

[8] T. Schmidt, K. Lischka, and W. Zulehner, Phys. Rev. B **45**, 8989 (1992).

[9] U. Kaufmann, M. Kunzer, M. Maier, H. Obloh, A. Ramakrishnan, B. Santic, and P. Schlotter, Appl. Phys. Lett. **72**, 1326 (1998).

[10] S. Assali, J. Greil, I. Zardo, A. Belabbes, M. W. A. De Moor, S. Koelling, P. M. Koenraad, F. Bechstedt, E. P. A. M. Bakkers, and J. E. M. Haverkort, J. Appl. Phys. **120**, (2016).

[11] E. Burstein, Phys. Rev. **93**, 632 (1954).

[12] M. Grundmann, *The Physics of Semiconductors* (Springer, 2006).

[13] I. Pelant and J. Valenta, *Luminescence Spectroscopy of Semiconductors* (2012).

[14] C. F. Klingshirn, *Semiconductor Optics* (Springer Berlin Heidelberg, Berlin, Heidelberg, 2012).

[15] E. F. Schubert, E. O. Göbel, Y. Horikoshi, K. Ploog, and H. J. Queisser, Phys. Rev. B **30**, 813 (1984).

[16] D. Ouadjaout and Y. Marfaing, Phys. Rev. B **41**, 12096 (1990).

[17] S. Assali, J. Nicolas, S. Mukherjee, A. Dijkstra, and O. Moutanabbir, Appl. Phys. Lett. **112**, 251903 (2018).